\documentclass[twocolumn, nofootinbib, aps, prl, floats, floatfix, amsmath, amssymb, longbibliography, superscriptaddress, preprintnumbers]{revtex4-1}
\usepackage{graphicx}
\usepackage{dcolumn}
\usepackage{bm}

\usepackage{lineno}

\usepackage{bbm}

\usepackage{amssymb}   
\usepackage{amsmath}   
\usepackage{appendix}
\usepackage{color}
\usepackage{xcolor} 
\usepackage{hyperref}
\usepackage[normalem]{ulem} 
\usepackage{orcidlink}
\def\bs{\mathbf{B}_{\rm S}}
\def\bi{\mathbf{B}_{\rm I}}

\begin{document}

\preprint{APS/123-QED}

\title{Particle Conversions Beyond the WKB Approximation and Solar-Induced Gravitational Waves from Dark Photon Dark Matter}

\author{Tengyu Ai}
\email{tengyuai@pku.edu.cn}
 \affiliation{School of Physics and State Key Laboratory of Nuclear Physics and Technology, Peking University, Beijing 100871, China}

\author{Yuxuan He}%
\email{heyx25@pku.edu.cn}
\affiliation{School of Physics and State Key Laboratory of Nuclear Physics and Technology, Peking University, Beijing 100871, China}

\author{Jia Liu}
\email{jialiu@pku.edu.cn}
\affiliation{School of Physics and State Key Laboratory of Nuclear Physics and Technology, Peking University, Beijing 100871, China}
\affiliation{Center for High Energy Physics, Peking University, Beijing 100871, China}

\author{Xiaolin Ma}
\email{themapku@stu.pku.edu.cn}
\affiliation{School of Physics and State Key Laboratory of Nuclear Physics and Technology, Peking University, Beijing 100871, China}

\author{Xiao-Ping Wang}
\email{hcwangxiaoping@buaa.edu.cn}
\affiliation{School of Physics, Beihang University, Beijing 100083, China}
\affiliation{Beijing Key Laboratory of Advanced Nuclear Materials and Physics, Beihang University, Beijing 100191, China}

\date{\today}

\begin{abstract}
We investigate the conversion of kinetic mixing dark photon dark matter into gravitational waves within the magnetic field of the Sun. Our study reveals that the WKB approximation is invalid in this scenario. We derive an analytic solution  for the conversion probability with unitary evolution feature. This solution aligns in form with previous studies on photon-gravitational wave conversion. Interestingly, it is applicable in situations where the WKB approximation fails. We extend the unitary evolution solution to other conversion processes, such as axion-photon and dark photon-photon conversions. When the WKB approximation conditions are met, this solution reduces to the WKB result.
We compute the characteristic strain of gravitational waves resulting from dark photon conversion in the solar magnetic field, spanning frequencies from $10^{-5}$ Hz to $10^6$ Hz. Our findings indicate that the characteristic strain derived from the unitary evolution solution differs significantly from that of the WKB solution. 
The resulting strain signal is far below the sensitivity of current gravitational wave interferometers. Nevertheless, we have proposed an exotic gravitational wave source, which could be useful in non-minimal dark sector models.
\end{abstract}

\maketitle


\section{Introduction}

The discovery of gravitational waves (GW) has opened a new era for astrophysics, cosmology, and particle physics~\cite{LIGOScientific:2016aoc}. The information carried by GWs can provide valuable insights into the history of cosmic evolution~\cite{Mazumdar:2018dfl,Caprini:2015zlo,Caprini:2019egz,Morrissey:2012db} and the nature of dark matter~\cite{Guo:2022dre, Bertone:2019irm, Chao:2023lox}.

A widely adopted approach in new physics calculations, such as photon-axion conversion~\cite{Raffelt:1987im, Huang:2018lxq, Hook:2018iia, Dessert:2019sgw} and photon-dark photon conversion~\cite{Mirizzi:2009iz, An:2020jmf}, involves utilizing the wave equation to compute the transition probability between states. Starting with the Lagrangian, one can derive the equations of motion and interpret them as wave equations to elucidate the evolution of each component. Throughout this process, it is advantageous to treat the solution as a slowly varying plane wave, known as the WKB approximation. However, this approximation relies on specific conditions, which are typically satisfied but can be violated in certain instances. In this work, we examine a potential GW signal from dark photon dark matter (DPDM) passing through the magnetic field inside the Sun, where the WKB approximation loses validity due to the slow velocity of the incoming dark matter (DM).

In the calculation, we demonstrate that we can derive an analytic result with the help of integrating out the photon due to the large plasma mass within the Sun. The analytic solution can be seen as a unitary evolution of the wave function, similar to the results in Ref.~\cite{Domcke:2020yzq}, and we denote it as the unitary evolution solution. This solution is general and works without the WKB approximation condition. We examine the WKB approximation condition and identify two typical cases where it holds: relativistic particle conversions~\cite{Liu:2023mll} and resonant particle conversions~\cite{Hook:2018iia, An:2020jmf}. We extend the unitary evolution solution to axion-photon and dark photon-photon conversion processes, applicable when the WKB approximation condition does not hold. We demonstrate that the unitary evolution  solution reduces to the WKB solution under the WKB approximation conditions. Other attention has also been given to calculations without the WKB approximation. For example, in Ref.~\cite{McDonald:2023ohd}, the authors calculate axion-photon conversion in 3D while using a Boltzmann-like equation.

Previous studies have investigated the conversion between GWs and photons. As a direct consequence of general relativity, GWs can transform into photons in the presence of a magnetic field, a phenomenon known as the Gertsenshtein effect~\cite{Gertsenshtein:1962kfm, Boccaletti:1970pxw, DeLogi:1977qe, Raffelt:1987im}. We incorporate the kinetic mixing dark photon~\cite{Fabbrichesi:2020wbt} into our model, allowing for conversions among these three components. 
GWs and dark photons can interconvert in the presence of a ``dark magnetic field," induced in our case by the magnetic field inside the Sun. Therefore, there can be GW production in the solar system when dark photon dark matter passes through the Sun. We calculate the corresponding dark magnetic field for various dark photon masses and estimate the final characteristic strain of GWs. Unfortunately, this resulting GW strain is too small to be detected with existing gravitational wave detectors. Nonetheless, we present a new exotic source of gravitational waves, especially at low frequencies, from the dark photon dark matter conversion. It should be noted that the dark sector may contain dark fermions which could generate the dark magnetic field directly. In such a case, the $A' \to \text{GW}$ conversion could occur without suppression from $\epsilon$, significantly enhancing the likelihood of such a conversion.

The outline of this paper is as follows.
In Sec.~\ref{sec:dp},  we provide a brief overview of the kinetic mixing dark photon model and specify its three different bases. We adopt the flavor basis for our calculations, with an equivalent calculation in the mass basis presented in the appendix.
In Sec.~\ref{sec:main}, we derive the equations of motion and solve for the conversion probability in the limit of a large plasma mass. We develop the general unitary evolution solution for dark photon and gravitational wave  conversion, and obtain the conversion probability.
In Sec.~\ref{sec:otherconversions}, we extend the unitary evolution solution to axion-photon and dark photon-photon conversion processes and demonstrate how it reduces to the WKB solution under appropriate conditions.
In Sec.~\ref{sec:strain}, we calculate the ``dark magnetic field" inside the Sun and determine the final characteristic strain of the resulting GWs. We also compare our results with those obtained using the WKB approximation, highlighting significant differences.
In Sec.~\ref{sec:conclusion}, we conclude.

\section{Dark Photon Lagrangians in Different Basis}
\label{sec:dp}

We can express the dark photon Lagrangian in different bases \cite{Fedderke:2021aqo}. The three primary choices of basis are the vacuum basis, the flavor basis, and the kinetic mixing basis. In this section, we will explore the relationships among these bases. It is important to note that the physical results are independent of the choice of basis. As a consistency check, we demonstrate in Appendix \ref{sec:der-EOM-massbasis} that the derived equation of motions (EOMs), and consequently the results, are indeed basis-independent.

The kinetic mixing basis explicitly represents the ``vector portal" nature of the mixing, which is the most commonly used framework~\cite{Caputo:2021eaa}. The massive dark photon interacts with the SM sector particles through the kinetic mixing to the SM photon. In this \textbf{kinetic mixing basis}, the effective Lagrangian for dark photon is given by 
\begin{align}
   \mathcal{L}_{\rm kin}&=-\frac{1}{4} F^{'}_{{\rm k},{\mu\nu}} F_{\rm k}^{'\mu\nu}-\frac{1}{4} F_{{\rm k}, {\mu\nu}} F_{\rm k}^{\mu\nu}+\frac{1}{2}m_A'^2 A^{'}_{{\rm k},\mu} A_{\rm k}^{' \mu}\nonumber \\
   &\quad -\frac{\epsilon}{2} F_{{\rm k},{\mu\nu}} F_{\rm k}^{'\mu\nu}+ A_{\rm k,\mu} J_{\rm EM}^{\mu},
\end{align}
where $F_{k,\mu\nu} \equiv \partial_\mu A_{k,\nu}-\partial_\nu A_{k,\mu}$ and $F_{k,\mu\nu}^{'}$ are the field strength tensors for the SM photon and the dark photon, respectively. The subscript $k$ denotes for the kinetic mixing basis, $\epsilon $ is the mixing coefficient, and $J_{\rm EM}$ is the electromagnetic current in the SM.

We can transfer from the kinetic mixing basis to the flavor basis via
\begin{align}
\begin{pmatrix}
    A_{\rm k}^\mu \\
    A_{\rm k}^{'\mu}
\end{pmatrix}
= \frac{1}{\sqrt{1+\epsilon^2}}
\begin{pmatrix}
    1 & 0 \\
    -\epsilon & 1
\end{pmatrix}
\begin{pmatrix}
    A_{\rm I}^\mu \\
    A_{\rm S}^\mu
\end{pmatrix}
\end{align}

where in the second equality, the leading order result in $\mathcal{O}(\epsilon)$ is kept.
Then, we can eliminate the kinetic mixing term and obtain the Lagrangian in the \textbf{flavor basis}. To leading order in $\epsilon$, it is given by:
\begin{align} 
   \mathcal{L}_{\rm flavor}&=-\frac{1}{4}F_{\rm I, \mu\nu}F_{\rm I}^{\mu\nu}-\frac{1}{4}F_{\rm S,\mu\nu}F_{\rm S}^{\mu\nu}+\frac{1}{2}m_A'^2A_{\rm S,\mu}A_{\rm S}^\mu \nonumber \\
   &\quad -\epsilon m_A'^2A_{\rm S,\mu}A_{\rm I}^\mu+A_{\rm I,\mu} J_{\rm EM}^{\mu},
   \label{Lag-FB}
\end{align}
where the subscript ``${\rm S(I)}$" represent ``sterile (interactive)" states, respectively. The detailed form of the Lagrangian is given in Appendix~\ref{sec:der-EOM-massbasis}. Importantly, only the interactive state, primarily coming from the photon state, directly couples to the SM EM current $J_{\rm EM}$, while the sterile state, predominantly originating from the dark photon state, does not interact with the EM current. In our study, we choose to work in the flavor basis because it is convenient to  analyze the dark photon-GW conversion and determine the dark photon magnetic field. 

The conversion between the interactive and sterile photons occurs solely through the mass mixing term. The presence of this mass mixing term indicates that the flavor eigenstates $A_{\rm S}$ and $A_{\rm I}$ are not the propagation eigenstates in the vacuum. In fact, we can obtain the vacuum basis by redefining the flavor basis:

\begin{align}
\begin{pmatrix}
A^\mu\\
A'^\mu
\end{pmatrix}=\frac{1}{\sqrt{1+\epsilon^2}}
\begin{pmatrix}
1 & ~\epsilon\\
-\epsilon & ~1
\end{pmatrix}
\begin{pmatrix}
A_{\rm I}^\mu\\
A_{\rm S}^\mu
\end{pmatrix}.
\end{align}
Then the Lagrangian in the \textbf{mass basis} (vacuum basis) to the leading order in $\mathcal{O}(\epsilon)$ can be rewritten  as
\begin{align}
   \mathcal{L}_{\rm vac}=&-\frac{1}{4}F_{\mu\nu}F^{\mu\nu}-\frac{1}{4}F'_{\mu\nu}F'^{\mu\nu}+\frac{1}{2}m_A'^2A'_\mu A'^\mu\nonumber\\
   &+(A_\mu-\epsilon A'_\mu)J_{\rm EM}^{\mu}.
   \label{eq:massbasis}
\end{align}
In this basis, the $A,~A'$ are the mass eigenstate in the vacuum. The dark photon dark matter is comprised of the mass eigenstate $A'$. In this basis the observed electromagnetic field is given by $\vec{B}_{\rm obs}=\vec{\nabla}\times (\vec{A}_\mu-\epsilon \vec{A'}_\mu)$~\cite{Dubovsky:2015cca}.

\section{The Dark Photon and Gravitational Wave Conversion}
\label{sec:main}
The well-known Gertsenshtein effect \cite{Gertsenshtein:1962kfm, Boccaletti:1970pxw, DeLogi:1977qe, Raffelt:1987im} describes the conversion of photons into gravitational wave in the presence of a strong magnetic field. Similarly, dark photons could undergo a comparable conversion process. In the context of Dark Photon Dark Matter (DPDM), it is intriguing to investigate the probability of DPDM converting into gravitational wave. Below, we derive the coupled equations of motion for photons, dark photons, and gravitational wave.

In our study, we start our calculations from the \textbf{Flavor} basis Lagrangian in Eq.~\eqref{Lag-FB}. We adopt the notation $\eta_{\mu\nu}={\rm diag}(+,-,-,-)$ and the metric fluctuation $h_{\mu\nu}\equiv g_{\mu\nu}-\eta_{\mu\nu}$ represents the classical field of the gravitational wave. The electromagnetic tensor under non-zero magnetic field $F_{\mu\nu}^{\rm ext}$ is~\cite{Domcke:2020yzq}:
\begin{align}
 F^{\mu\nu}&\supseteq \left(\eta^{\mu\lambda} \partial_{\lambda} A^{\nu}-\eta^{\nu\lambda} \partial_{\lambda} A^{\mu}\right)\nonumber \\
 &\quad + \left(\eta^{\alpha\mu}\eta^{v\nu}-h^{\alpha\mu}\eta^{\beta\nu}-\eta^{\alpha\mu}h^{\beta\nu} \right) F^{\text{ext}}_{\alpha\beta}\,,
\label{eq:F}
\end{align}
where we keep terms up to linear order in $h$ or $A$.
We choose the Lorentz gauge for photon and dark photon and harmonic gauge for gravitational wave, 

\begin{align}
\partial_\nu A_{\rm I/S}^\nu =0 ,~~~~~~
\partial_{\nu}h^{\beta\nu} = \frac{1}{2}\eta^{\beta \nu}\partial_{\nu}(\eta^{\rho \sigma}h_{\rho \sigma}).
\end{align}

To illustrate the interaction between gravitational wave, photons, and dark photons, one can examine the equations of motion in curved space-time by expanding the space-time metric to leading order. In curved space-time, the Euler-Lagrange equation takes the form:
\begin{equation}
  \nabla_\mu\frac{\partial \mathcal{L}}{\partial (\partial_\mu \phi)}=\frac{\partial \mathcal{L}}{\partial \phi},
\end{equation}
where $\nabla$ is the covariant derivative of spacetime. The EOMs of photons and dark photons thus become
\begin{equation}
\begin{cases}
&\nabla_\mu F_{\rm I}^{\mu\nu}-\epsilon m_{A'}^2A_{\rm S}^{\nu}+\Delta_{\rm I}A_{\rm I}^{\nu}+J_{\rm EM,ext}^\nu=0,   \\
&\nabla_\mu F_{\rm S}^{\mu\nu}-\epsilon m_{A'}^2A_{\rm I}^{\nu}+m_{A'}^2A_{\rm S}^{\nu}=0, 
\end{cases}
\label{eq:dp}
\end{equation}
where $\displaystyle\nabla_{\nu}F^{\mu \nu}=\frac{1}{\sqrt{-g}}\partial_{\nu}(\sqrt{-g}F^{\mu \nu})$ and $J_{\rm EM,ext}$ is the external EM background current. We define 
\begin{align}
\Delta_{\rm I}\equiv\omega^2_{\rm pl}+\Delta_{\rm vac}=\frac{e^2n_e}{m_e}-\frac{28\omega^2\alpha^2B_{\rm obs}^2}{45m_e^4},
\end{align}
explicitly include the plasma effect for the interactive field $A_{\rm I}$ with the plasma mass $\omega_{\rm pl}^2=e^2n_e/m_e$ and QED vacuum polarization effect $\Delta_{\rm vac}$~\cite{Liu:2023mll, Raffelt:1987im}. After applying the classical equation for the external field 
\begin{align}
\partial_\mu\eta^{\alpha\mu}\eta^{\beta\nu} F^{\text{ext}}_{\alpha\beta}+J_{\rm EM,ext}^\nu=0,
\end{align}
the EOMs of photon and DP Eq.~\eqref{eq:dp} become
\begin{equation}
\left(\Box+\omega^2_{\rm p} \right)\begin{pmatrix}A_{\rm I}^1 \\ A_{\rm I}^2\end{pmatrix}= 
\begin{pmatrix}
\partial_{l} h_+ & \partial_{l} h_\times\\
\partial_{l} h_\times & -\partial_{l} h_+
\end{pmatrix}
\begin{pmatrix}
B_{\rm I}^2\\
-B_{\rm I}^1\\
\end{pmatrix}
+\epsilon m_{A'}^2\begin{pmatrix}A_{\rm S}^1 \\ A_{\rm S}^2\end{pmatrix}\,,
\end{equation}
\begin{equation}
\left(\Box+m_{A'}^2 \right)\begin{pmatrix}A_{\rm S}^1 \\ A_{\rm S}^2\end{pmatrix} = 
\begin{pmatrix}
\partial_{l} h_+ & \partial_{l} h_\times\\
\partial_{l} h_\times & -\partial_{l} h_+
\end{pmatrix}
\begin{pmatrix}
B_{\rm S}^2\\
-B_{\rm S}^1\\
\end{pmatrix}
+\epsilon m_{A'}^2\begin{pmatrix}A_{\rm I}^1 \\ A_{\rm I}^2\end{pmatrix}\,,
\end{equation}
where $\Box=\partial_0^2-\partial_i^2$, $\partial_i = (0,0,\partial_{l})$, with the third component along the propagation direction. The superscripts $1$ and $2$ denote the two orthogonal directions in the transverse plane of the propagation direction. Additionally, $F^\text{ext}_{23} =-B^1$ and $F^\text{ext}_{13} =B^2$. It is worth noting that the observed external magnetic field $\vec{B}_{\rm obs}$ is the same as $\vec{B}_{\rm I}$.

In order to solve the wave function, we rewrite the Einstein equation under the aforementioned gauge choice as 
\begin{align}
   \Box h_{\mu\nu}=-\kappa^2T_{\mu\nu},~~~~\text{with}~\kappa\equiv\sqrt{16\pi G}.
\end{align}
The total Maxwell stress tensor with the absence of the $\vec{E}$ field for the photon and DP are given as 
\begin{align}
      T_{ij}& = (B_{\rm I}^{\rm tot})^i(B_{\rm I}^{\rm tot})^j- \frac{1}{2}\delta^{ij}(B_{\rm I}^{\rm tot})^2
      +(B_{\rm S}^{\rm tot})^i(B_{\rm S}^{\rm tot})^j\nonumber\\
      &\quad - \frac{1}{2}\delta^{ij}(B_{\rm S}^{\rm tot})^2 .
\end{align}
where we include the both contribution of the external field and the dynamic fields. We retain only the term linear in $A_{\rm I/S}$ and drop higher order i$A_{\rm I/S}$ terms and the term quadratic in the external field as they do not contribute to the generation of gravitational waves.

\begin{figure}
\centering
\includegraphics[width= 0.99 \linewidth]{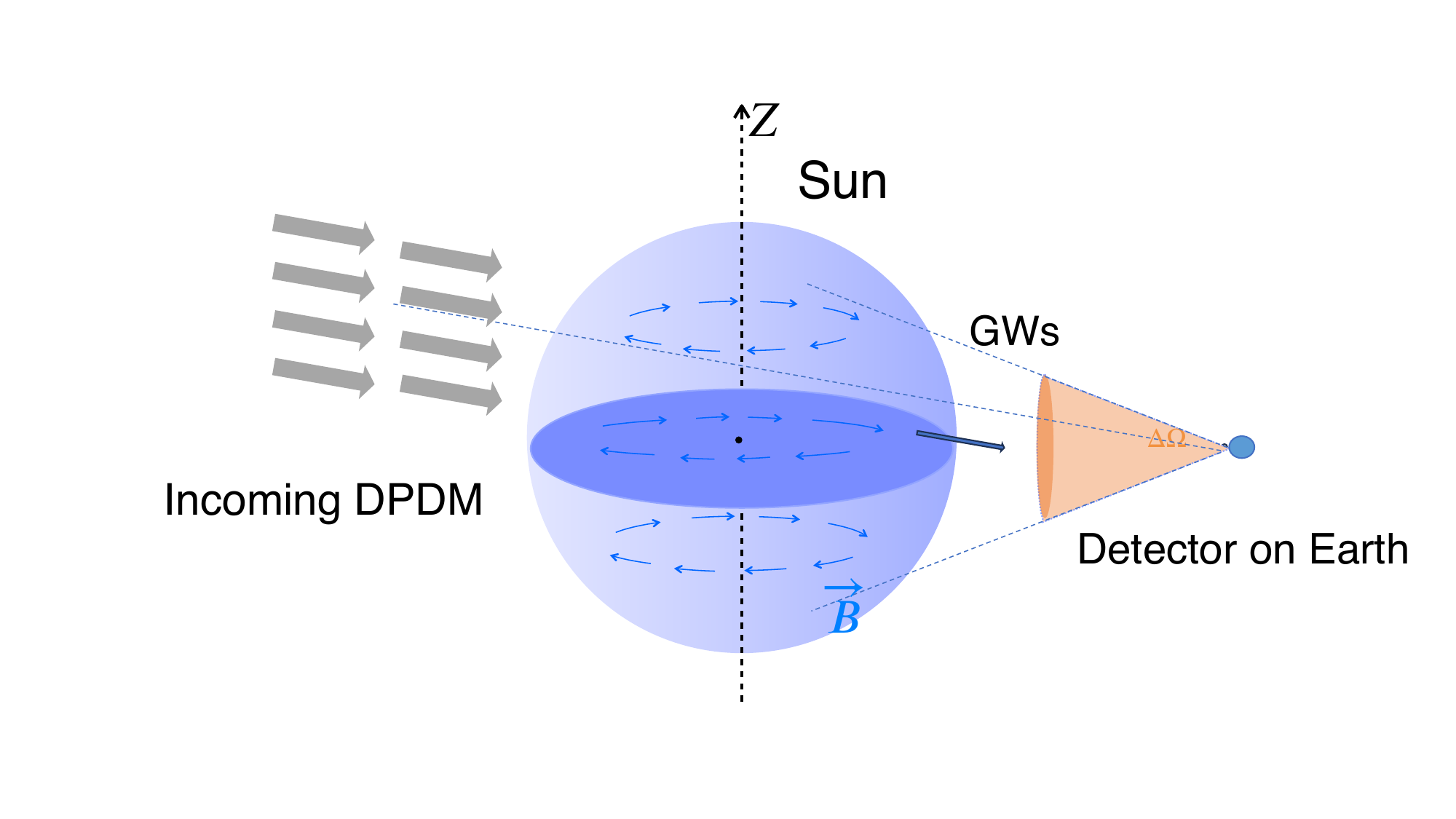}
\includegraphics[width= 0.75\linewidth]{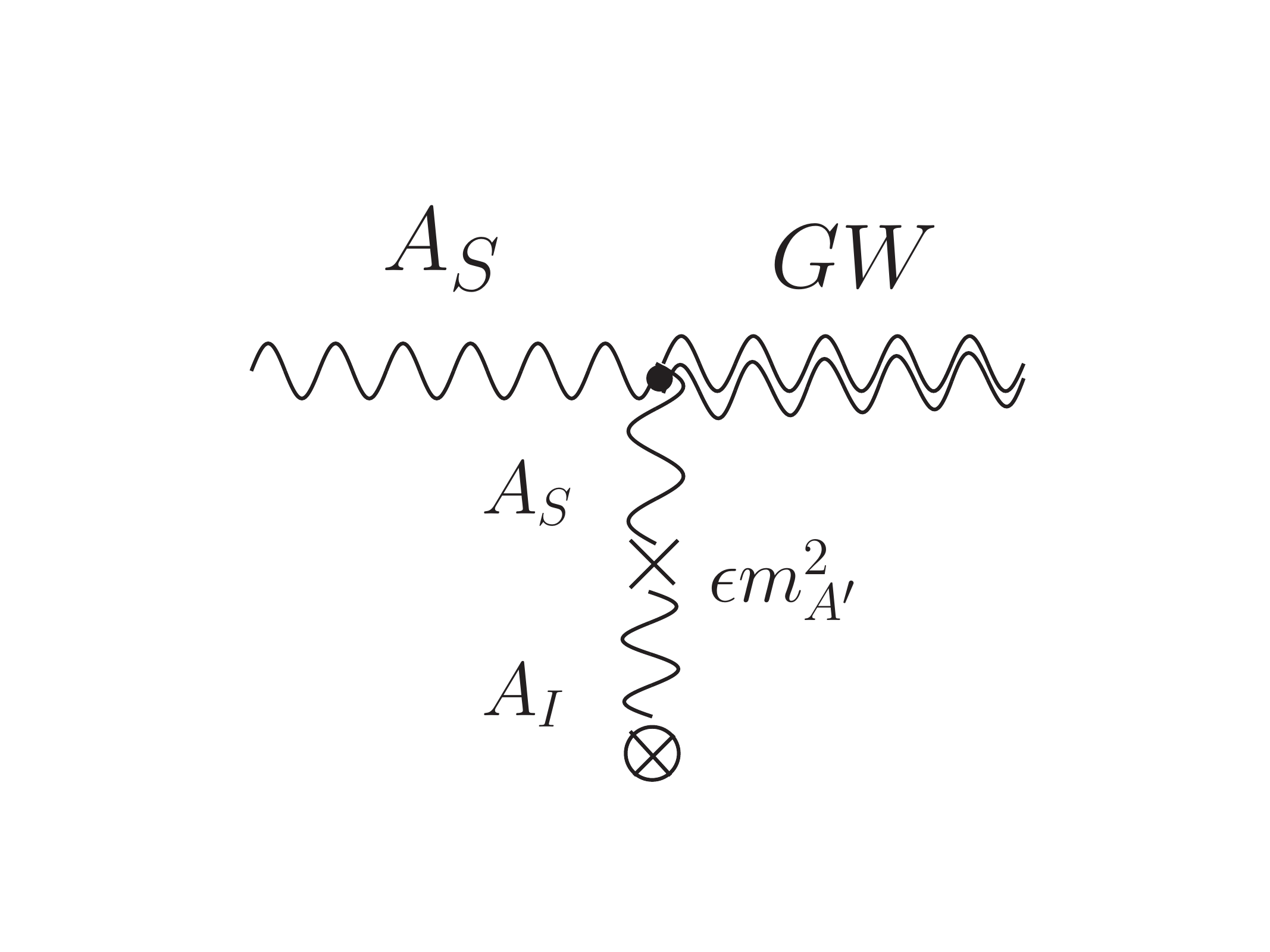}
\caption{
The illustrative picture and relevant Feynman diagram of conversion of DPDM into gravitational waves within the solar magnetic field. The Sun's self-rotation axis is tilted by about 7.25 degrees from the axis of the Earth’s orbit~\cite{sunselfrot}, which has been neglected here. In flavor basis the active magnetic field $\bi$ of the Sun could source dark magnetic field $\bs$ via quadratic mass mixing terms and thus could mediate the conversion of DPDM into gravitational waves.  }
\label{fig:illu} 
\end{figure} 

We can simplify the coupled EOMs for the photon, DP, and graviton as follows: 
\begin{align}
\begin{cases}
(\Box + \Delta_{\rm I})A_{\rm I}^\lambda+\kappa {\rm B}_{\rm I}\partial_l \tilde h^{\lambda}-\epsilon m_{A'}^2 A_{\rm S}^\lambda=0,\\
(\Box + m_{A'}^2)A_{\rm S}^\lambda+\kappa {\rm B}_{\rm S}\partial_l \tilde h^{\lambda}-\epsilon m_{A'}^2 A_{\rm I}^\lambda=0,\\
\Box \tilde h^\lambda-\kappa {\rm B}_{\rm S}\partial_l A_{\rm S}^{\lambda}-\kappa B_{\rm I} \partial_l A_{\rm I}^\lambda=0,
\end{cases}
\label{eq:full-eoms}
\end{align}
where index $\lambda=1, 2$ for $A_{\rm I/S}$, $\lambda=\times,+$ for $\tilde h$ and $h\equiv\kappa \tilde{h}$.

To solve the coupled equations of motion, one usually assume radial plane wave solutions $A_{\rm I}(r,t)=e^{i(\omega t-k r)}\tilde{A}_{\rm I}(r)$, $A_{\rm S}(r,t)=e^{i(\omega t-k r)}\tilde{A}_{\rm S}(r)$ and $\tilde{h}(r,t)=e^{i(\omega t-k r)}\tilde{\tilde{h}}(r)$ with dark photon dark matter on-shell condition $\omega^2=k^2+m_{A'}^2$. We simplify the second-order differential equations to first-order differential equations using the WKB approximation, as in Refs.~\cite{Raffelt:1987im, Boccaletti:1970pxw, An:2020jmf, Guarini:2020hps, Gu:2023fqm}. However, as discussed in Sec.~\ref{sec:wkb-approx}, the WKB approximation requires $|\tilde{A}_{\rm I/S}''(r)| \ll k|\tilde{A}_{\rm I/S}'(r)|$.  Unfortunately, in the case of dark photon dark matter conversion into gravitational waves, this condition is \textit{not} satisfied. Therefore, a direct solution to the coupled second-order differential equations is required.

\subsection{Decoupling of Photon States Inside the Sun}
\label{sec:decoupling-photons}
The main focus of this paper is the DPDM-GW conversion inside the Sun. The electron number density in the Sun (at half the Sun's radius) is about $1 \, \text{mol} / \text{cm}^3$~\cite{Couvidat:2002gvk}, which results in a plasma frequency of $\omega_{\text{pl}} \approx 10 \, \text{eV} \approx 10^4 \, \text{THz}$. Therefore, if we consider DP masses ranging from $10^{-5}$ to $10^3 \, \text{Hz}$ for the interest of GW experiments, the DPDM mass of interest typically satisfies $m^2_{A^\prime} \ll \Delta_\text{I}$. This allows us to integrate out the interacting state $A_\text{I}$ (mostly the normal photon state). Specifically, under the condition $\Box A_\text{I} \ll \Delta_\text{I} A_\text{I} $, we can solve the first equation as:
\begin{equation*}
    A_{\rm I}^\lambda = \frac{-1}{\Delta_{\rm I}}\left(\kappa {\rm B}_{\rm I}\partial_l \tilde h^{\lambda}-\epsilon m_{A'}^2 A_{\rm S}^\lambda \right),
\end{equation*}
then substitute it into the second and third lines in Eq.~\eqref{eq:full-eoms}, yielding:
\begin{align}
\begin{cases}
(\Box + m_{A'}^2)A_{\rm S}^\lambda+\kappa {\rm B}_{\rm S}\partial_l \tilde h^{\lambda} +\displaystyle\frac{\epsilon m_{A'}^2}{\Delta_{\rm I}}\left(\kappa {\rm B}_{\rm I}\partial_l \tilde h^{\lambda}-\epsilon m_{A'}^2 A_{\rm S}^\lambda \right)=0,\\
\Box \tilde h^\lambda-\kappa {\rm B}_{\rm S}\partial_l A_{\rm S}^{\lambda}+\displaystyle\frac{\kappa B_{\rm I}}{\Delta_{\rm I}} \partial_l  \left(\kappa {\rm B}_{\rm I}\partial_l \tilde h^{\lambda}-\epsilon m_{A'}^2 A_{\rm S}^\lambda \right)=0.
\end{cases}
\label{eq:reduced-2x2}
\end{align}
Since the $\Delta_\text{I}$ term is very large, we can omit the terms that are suppressed by $1/\Delta_\text{I}$. Consequently, the EOMs become
\begin{align}
\begin{cases}
(\Box + m_{A'}^2)A_{\rm S}^\lambda+\kappa {\rm B}_{\rm S}\partial_l \tilde h^{\lambda} =0,\\
\Box \tilde h^\lambda-\kappa {\rm B}_{\rm S}\partial_l A_{\rm S}^{\lambda}=0,
\end{cases}
\label{eq:sim-eoms}
\end{align}
where we revert to the dark photon and gravitational wave version of the Gertsenshtein equations.
One can also interpret the decoupling in another way. For an incoming non-relativistic DPDM with energy 
$\omega \sim m_{A'} \ll \omega_{\rm pl}$, the momentum of the converted photon $k_1$ will become imaginary,  $k_1 \approx i \omega_{\rm pl}$. As a result, the converted photon will be damped immediately within a distance of about $\omega_{\rm pl}^{-1}$.

\subsection{General Solutions of Second-Order Differential EOMs
}

In the previous subsection, we listed the EOMs for photons, dark photons, and gravitational waves, which describe the conversion between their amplitudes in the presence of external fields. We will study the DPDM conversion into gravitational waves due to the presence of an external magnetic field inside the Sun. As we mentioned, the WKB approximation does not work in this case. Therefore, we will provide the direct solution to the second-order differential EOMs in this section.

Although we discussed in the last section that the photon state can be integrated out inside the Sun, resulting in reduced EOMs for dark photon and gravitational wave states, the direct solution can be easily generalized to include three states: photon, dark photon, and gravitational wave. Thus, we present the general solution for the three-state conversion process, which may benefit interested readers.

We first start the calculation in homogeneous external magnetic field, and then promote the calculation results to general inhomogeneous magnetic field case. For a given particle with propagation energy $\omega$, only certain $k$-modes whose EOM determinant equals to zero are allowed to propagate. Substituting $\partial_t \to -i \omega$ and ignoring the common phase $e^{-i \omega (t - t_0)}$, we rewrite the EOMs ~\eqref{eq:full-eoms} under the following ansatz:

\begin{align}
 \hat{f}\cdot  \begin{pmatrix}
        A_I(r) \\
        A_S(r)  \\
        \tilde{h}(r)
    \end{pmatrix} = \hat{f}\cdot e^{i \mathbf{K}(r-r_0)} \psi(t_0,r_0)= \vec{0},
    \label{eq:eom1}
\end{align}
with the differential operator 
\begin{align}
    \hat{f}\equiv 
     \begin{pmatrix}
        -\omega^2-\partial_r^2 +\Delta_{\rm I}^2 & -\epsilon m_{A'}^2 &  B_{\rm I}\kappa \partial_r\\
        -\epsilon m_{A'}^2  & -\omega^2-\partial_r^2 +m_{A'}^2 &  B_{\rm S}\kappa \partial_r\\
        -B_{\rm I}\kappa \partial_r & -B_{\rm S}\kappa \partial_r & -\omega^2-\partial_r^2
    \end{pmatrix},
\end{align}
and $\psi(t_0,r_0)\equiv \left\{A_{\rm I}(t_0,r_0),A_{\rm S}(t_0,r_0),\tilde{h}(t_0,r_0) \right\}^{\rm T}$ is state vector of the interacting field, sterile field and gravitational wave field at the initial point $(t_0,r_0)$.  

The phase factor $e^{i \mathbf{K}(r-r_0)}$ encodes the unitary evolution of three fields with $\mathbf{K}$ being a Hermitian matrix. 
To determinate the exact form of $\mathbf{K}$ and solve the differential equation, we first perform a Fourier transformation to rewrite the EOMs in momentum space. Then, we obtain $f(k)$ by substituting $\partial_r \to i k$ in $\hat{f}$, 
\begin{align}
    f(k)\equiv 
     \begin{pmatrix}
        -\omega^2+k^2 +\Delta_{\rm I}^2 & -\epsilon m_{A'}^2 &  i B_{\rm I}\kappa k\\
        -\epsilon m_{A'}^2  & -\omega^2+k^2 +m_{A'}^2 & i B_{\rm S}\kappa k\\
        -i B_{\rm I}\kappa k & -i B_{\rm S}\kappa k & -\omega^2+k^2
    \end{pmatrix}.
\end{align}

The next step is to require that the determinant of $f(k)$ equals zero,
\begin{align}
    {\rm Det} \left[  f(k) \right]= 0 \,,
\end{align}
which allows us to obtain three forward propagation values, $k_{i = 1, 2, 3}$. For each fixed $k_i$, the $3 \times 3$ matrix $f(k_i)$ has a zero determinant, meaning it should have a normalized vector $\nu_i$, each with $\nu_i^\dagger \cdot \nu_i = 1$, and has zero eigenvalues. Explicitly, for each fixed $i$, it satisfies the following conditions,
\begin{align}
    f(k_i) \cdot \nu_i = \vec{0} \quad \text{and} \quad \hat{f} \cdot e^{i k_i (r-r_0) } \nu_i = \vec{0}\,.
    \label{eq:EOM-solution-non-orthogonal}
\end{align}

Note that there is a significant difference from the standard procedure of solving the eigensystem of a Hermitian matrix $A$, which involves setting the determinant of $- \omega^2 \mathbf{I} + A$ to zero and obtaining three solutions for $\omega$. In the standard procedure, the three eigenvalues are automatically orthogonal to each other. In our case, we are solving for $k$, where the off-diagonal terms of $f(k)$ contain $k$, to satisfy the determinant of $f(k)$ equals zero. Therefore, the $\nu_i$ are the eigenvectors corresponding to the zero eigenvalue for three different matrices $f(k_i)$. As a result, the three normalized $\nu_i$s are not necessarily orthogonal with each other.

Nevertheless, we can arrive at the general solution of the second-order differential EOMs (Eq.~\eqref{eq:eom1}) by linearly combining the three independent solutions as follows:
\begin{align}
    \begin{pmatrix}
        A_I(r) \\
        A_S(r)  \\
        \tilde{h}(r)
    \end{pmatrix} = \sum_{j} \alpha_j e^{i k_j (r-r_0)} \nu_j \,,
    \label{eq:generalsolution-3states}
\end{align}
 where $\alpha_j$ are constants to be determined by the initial conditions $A_I(r_0)$, $A_S(r_0)$, and $h(r_0)$. The above solution describes the field amplitude evolution with distance $r$ in constant external magnetic fields $B_I$ and $B_S$. For varying external fields, one can evolve the field amplitude in small steps to achieve the final amplitudes.

\subsection{The Unitary Evolution Solution}
\label{sec:unitary-evolution-solution}

In this subsection, we aim to derive the unitary evolution operator solution, which is analogous to quantum mechanics and thus easier to understand for the conversion probability. Moreover, this solution is similar to the form in Refs.~\cite{Gertsenshtein:1962kfm, Boccaletti:1970pxw, DeLogi:1977qe, Raffelt:1987im}. In this form, it is convenient to compare the solutions with and without the WKB approximation.

The difficulty in deriving a unitary evolution operator solution is that the three normalized $\nu_i$ are \textit{not} orthogonal. We will show that for two states in the flavor basis, one can always achieve the unitary evolution operator solution regardless of the form of $\hat{f}$. Fortunately, in Section \ref{sec:decoupling-photons}, we have already shown that we can decouple the photons and keep only the dark photon and gravitational wave states, which reduces the $3 \times 3$ matrix to a $2 \times 2$ matrix. The new differential operator of the EOMs becomes
\begin{align}
    \hat{f}_{2\times 2} =      \begin{pmatrix}
         -\omega^2-\partial_r^2 +m_{A'}^2 &  B_{\rm S}\kappa \partial_r\\
        -B_{\rm S}\kappa \partial_r & -\omega^2-\partial_r^2
    \end{pmatrix},
    \label{eq:f2x2}
\end{align}
which matches Eq.~\eqref{eq:reduced-2x2}. The unitary evolution operator has the form 
\begin{align}
e^{i \mathbf{ K}(r-r_0)}&\equiv \begin{pmatrix} \bar{\nu}_2 &\bar{\nu}_3\end{pmatrix} 
\begin{pmatrix} 
e^{ik_2(r-r_0)}& 0\\
 0& e^{ik_3(r-r_0)}\\
\end{pmatrix} 
\begin{pmatrix}  \bar{\nu}_2^\dagger \\ \bar{\nu}_3^\dagger\end{pmatrix} \nonumber\\
&= \bar{V} \cdot e^{i D (r-r_0)}\cdot \bar{V}^{\dagger},
\end{align}
where $\bar{V} \equiv \begin{pmatrix} \bar{\nu}_2 & \bar{\nu}_3 \end{pmatrix}$ is a unitary matrix, satisfying $\bar{V} \bar{V}^\dagger = I$, and $D \equiv {\rm diag}(k_2, k_3)$ is a diagonal matrix with $k_{2,3}$ satisfying ${\rm Det}[f_{2\times 2}(k)] = 0$. 

Therefore, the Hermitian matrix $\mathbf{K}$ can be expressed as $\mathbf{K} = \bar{V} \cdot D \cdot \bar{V}^\dagger$, and the phase factor $e^{i \mathbf{K} (r - r_0)}$ is the unitary evolution operator, which satisfies the solution $\hat{f}_{2 \times 2} \cdot e^{i \mathbf{K} (r - r_0)} \psi(t_0, r_0 ) = \vec{0}$.

Next, we would like to find a way to transform the normalized non-orthogonal vectors $\nu_{2,3}$ into normalized orthogonal vectors $\bar{\nu}_{2,3}$. Following Eq.~\eqref{eq:EOM-solution-non-orthogonal} in the previous section, we can rewrite it as a matrix form as:
\begin{align}
    &\hat{f} \cdot \begin{pmatrix} \nu_2 &\nu_3\end{pmatrix} 
\begin{pmatrix} 
e^{ik_2(r-r_0)}& 0\\
 0& e^{ik_3(r-r_0)}\\
\end{pmatrix} \nonumber\\
&= \hat{f} \cdot V \cdot e^{i D (r-r_0)} = 0_{2\times 2},
\label{eq:2x2-fhat-solution}
\end{align}
where \( V \equiv \begin{pmatrix} \nu_2 & \nu_3 \end{pmatrix} \). We can use two diagonal matrices, \( Y = {\rm diag}(y_2, y_3) \) for rescaling the field amplitudes and \( Z = {\rm diag}(z_2, z_3) \) for normalizing each independent solution. We redefine the field strength \(\psi(r) = (A_{\rm I}(r),\tilde{h}(r))^{\rm T}\) as:
\begin{align}
   \bar{\psi}(r) =  \begin{pmatrix}  \bar{A}_{\rm I}(r) \\ \bar{\tilde{h}}(r) \end{pmatrix} = \begin{pmatrix}  y_2 A_{\rm I}(r) \\ y_3 \tilde{h}(r) \end{pmatrix} =
   Y \cdot \begin{pmatrix}  A_{\rm I}(r) \\ \tilde{h}(r) \end{pmatrix}   = Y \cdot \psi(r).
\end{align}
Therefore, the differential EOMs transform accordingly from \(\hat{f} \psi(r) = 0\) to \(\bar{f} \bar{\psi}(r) = 0\) with \(\bar{\psi}(r) \equiv (\bar{A}_I(r), \bar{h}(r))^T\), as long as the operator transforms as follows:
\begin{align}
    \hat{f } = Y^{-1} \cdot \bar{f} \cdot Y.
\end{align}
Multiplying \(Z\) on the right-hand side of Eq.~\eqref{eq:2x2-fhat-solution} and changing \(\hat{f}\) to \(\bar{f}\), we have
\begin{align}
    0_{2\times 2} &= \hat{f} \cdot V \cdot e^{i D (r-r_0)} \cdot Z \nonumber\\
    & = Y^{-1} \cdot \bar{f} \cdot (Y V Z) \cdot e^{i D (r-r_0)} = Y^{-1} \cdot \bar{f} \cdot \bar{V} \cdot e^{i D (r-r_0)}.
\end{align}
In the second equality, we have used the commutation of \(e^{i D (r - r_0)}\) and \(Z\) because both are diagonal matrices. In the third equality, we can choose an appropriate \(Y\) to make \(Y \cdot \nu_2\) and \(Y \cdot \nu_3\) orthogonal and choose an appropriate \(Z\) to make the two orthogonal vectors \(z_2 Y \cdot \nu_2\) and \(z_3 Y \cdot \nu_3\) normalized. Therefore, we have obtained the unitary matrix
\begin{align}
    \bar{V} = Y V Z  = \begin{pmatrix} \bar{\nu}_2 & \bar{\nu}_3 \end{pmatrix}  \,.
\end{align}

In summary, we have approved that for a general differential operator $\hat{f}$ for EOMs without WKB approximation, we can rescale the field to the basis $\bar{\psi}(r)$. With the new operator $\bar{f}$ in the rescaled basis, we prove the unitary evolution solution is
\begin{align}
    \vec{0} = \bar{f} \cdot \bar{V} \cdot e^{i D (r-r_0)} \cdot \bar{V}^{\dagger} \bar{\psi}(r_0) = \bar{f} \cdot e^{i \mathbf{K}(r-r_0) } \bar{\psi}(r_0)
\end{align}

\subsection{The Conversion Probability}

In this section, we will calculate the conversion probability for dark photons and gravitational waves. Given the specific form of $\hat{f}_{2\times 2}$ in Eq.~\eqref{eq:f2x2}, we can calculate the eigenvalues and eigenvectors as follows:
\begin{equation}
\begin{aligned}
k_2=\frac{1}{2}\left(\sqrt{\omega^2(1+v)^2+B_{\rm S}^2\kappa^2}-\sqrt{\omega^2(1-v)^2+B_{\rm S}^2\kappa^2}\right), \\
k_3=\frac{1}{2}\left(\sqrt{\omega^2(1+v)^2+B_{\rm S}^2\kappa^2}+\sqrt{\omega^2(1-v)^2+B_{\rm S}^2\kappa^2}\right), \\
\bar{\nu}_2=z_2\left(\begin{array}{c}
-i \sqrt{v}B_{\rm S}\kappa k_2\\k_2^2-\omega^2v^2 \end{array}\right) \,, \quad 
\bar{\nu}_3=z_3 \left(\begin{array}{c}
-i\sqrt{v}B_{\rm S}\kappa k_3\\k_3^2-\omega^2v^2 \end{array}\right),
\end{aligned}
\end{equation}
where we use the incoming DP velocity $ v\equiv \sqrt{1 - m_{A'}^2 / \omega^2}$. 
The $z_{2,3}$ are the normalization factors for the eigenvectors from the matrix \(Z\).

To make the eigenvectors $\bar{\nu}_{2,3}$ orthogonal, the field vector has to be rescaled by the diagonal matrix \(Y\). The rescaling matrix \(Y\) and the rescaled basis are
\begin{equation}
 Y = \begin{pmatrix} 
\sqrt{v} & \quad 0\\
 0  & \quad 1\\
\end{pmatrix}  \,, \quad   \bar{\psi}( r_0) = Y \cdot \psi(r_0) =  \left(\begin{array}{c}
\sqrt{v} A_{\rm I}(r_0) \\  \tilde{h}(r_0)  \end{array}\right).
\end{equation}
The resulting expression for the Hermitian matrix \(\mathbf{K}\) is:
\begin{align}
    \mathbf{K} = \sum_{i =2,3} k_i \bar{\nu}_i \bar{\nu}_i^\dagger &=
    \begin{pmatrix}
     \frac{v \sqrt{B_{\rm S}^2 \kappa^2 + (v + 1)^2 \omega^2}}{v + 1} & -\frac{i \kappa B_{\rm S} \sqrt{v}}{v + 1} \\
     \frac{i \kappa B_{\rm S} \sqrt{v}}{v + 1} & \frac{\sqrt{B_{\rm S}^2 \kappa^2 + (v + 1)^2 \omega^2}}{v + 1}
    \end{pmatrix}  .
\end{align}
For the non-relativistic dark matter, we have
\begin{align}
\mathbf{K} 
\approx 
   \begin{pmatrix}
     \omega v & -i \kappa B_{\rm S} \sqrt{v} \\
     i \kappa B_{\rm S} \sqrt{v} & \omega 
    \end{pmatrix},
    \label{eq:homo_res_K}
\end{align}
with $v \ll 1$ (i.e. non-relativistic limit,) and $\omega \gg \kappa B_{\rm S}$. 

After the rescaling by a factor \( v \), one can check that the modulus of \( \bar{\psi} \) is proportional to the energy flux of DP and GW in the propagation direction, which is conserved during the propagating.


Consequently, the complete unitary evolution solution in a homogeneous background magnetic field is given by 
\begin{equation}
    \bar{\psi}(t,r) = e^{-i\omega (t - t_0)} e^{i {\mathbf K} (r - r_0)} \bar{\psi}(t_0, r_0) \equiv 
    e^{-i\omega (t - t_0)} U \bar{\psi}(t_0, r_0).
    \label{eq:homo_res}
\end{equation}

There are a few points to comment on for the solution.
Firstly, it is evident that the magnitude of \(\bar{\psi}\) is conserved because the evolution matrix $U = \exp{(i \mathbf{K} (r - r_0))}$  is locally unitary. Therefore, the off-diagonal term \(U_{12}\) can be interpreted as the conversion probability of a dark photon to a gravitational wave after traveling a distance \((r - r_0)\), following Ref.~\cite{Domcke:2020yzq}.

Secondly, the derived Hermitian matrix \(\mathbf{K}\) shares the same form as in Ref.~\cite{Domcke:2020yzq}, except for a different definition in \(\mu\). This similarity arises because the plasma frequency \(\omega_{\rm pl}\) only affects the active state \(A_{\rm I}\) in the flavor basis. For a very large plasma frequency \(\omega_{\rm pl}\), the active photon state \(A_{\rm I}\) effectively decouples from the other two states, reducing the problem to dark photon-gravitational wave conversion, which is similar to the photon-gravitational wave conversion in Ref.~\cite{Domcke:2020yzq}.

The unitary evolution operator $U$ can be explicitly written as
\begin{align}
U &= e^{i\mathbf{K}(r-r_0)} \nonumber \\
  &= e^{\frac{i}{2} \mathbf{Tr}[\mathbf{K}] (r-r_0)} \Bigg[ \cos\left(\frac{r-r_0}{r_{\rm osc}}\right) \cdot \mathbbm{1} \nonumber \\
  &\quad + i r_{\rm osc} \sin\left(\frac{r-r_0}{r_{\rm osc}}\right) \left(\mathbf{K} - \frac{1}{2} \mathbf{Tr}[\mathbf{K}] \mathbbm{1}\right) \Bigg],
\end{align}
where the oscillation length is
\begin{align}
    r_{\rm osc} \equiv 2/(k_3-k_2)=2\left(\sqrt{\omega^2(1 + v)^2+B_{\rm S}^2\kappa^2}\right)^{-1} \approx \frac{2}{ \omega } .
\end{align}
The transition probability for dark photon to gravitational wave in a homogeneous background is given by 
\begin{align}
P_{\rm homo}(A'\to \text{GW})&=\left|{U}_{21}(r,r_0)\right|^2 \nonumber \\
&= \left| \kappa B_{\rm S} \sqrt{v} \right|^2 r^2_{\rm osc}\sin^2\left(\frac{r-r_0}{r_{\rm osc}}\right).
\label{eq:homo_prob}
\end{align}
The physical conversion probability should be consistent between calculations performed in different bases. The above calculation is carried out in the flavor basis. We explicitly perform the same calculation in the mass basis and find the results in different bases align with each other. The detailed derivation for the mass basis is available in Appendix~\ref{sec:der-EOM-massbasis}.

Under realistic conditions, the magnetic fields and plasma densities can exhibit heterogeneity and complexity, necessitating an extension beyond simple homogeneous calculations. To address these inhomogeneities, we partition the region into small patches where the plasma density and magnetic field are uniform. Consequently, Eq.~\eqref{eq:homo_res} remains applicable. To ensure continuity, we smooth out the field values at the boundaries of adjacent patches. 
Connecting the solution of each patches, one can have the following solution for an inhomogeneous background in the form of first-order differential equation~\cite{Domcke:2020yzq}: 
\begin{align}
    &\bar{\psi}(t,r) = e^{-i\omega (t - t_0)} U(r,r_0)\bar{\psi}(t_0,r_0), \nonumber \\
    &\partial_r {U}(r,r_0) \approx i\mathbf K(r) {U}(r,r_0).
    \label{eq:inhomo_res}
\end{align}
The $\mathbf K(r)$ has the form in Eq.~\eqref{eq:homo_res_K}, but with magnetic fields $B_{\rm S}(r)$ values in this patch. 

In the flavor basis, the off-diagonal matrix elements of \(\mathbf{K}\) are much smaller compared to the diagonal elements, allowing us to treat the off-diagonal matrix elements as small perturbations. We can utilize perturbation methods and obtain the conversion probability in an inhomogeneous background as~\cite{Raffelt:1987im, Domcke:2020yzq}
\begin{widetext}
\begin{align}
P_{\rm inhomo}(A'\to \text{GW})&=\left|{U}_{21}(r,r_0)\right|^2 \approx \left|\int_{r_0}^rdr'e^{-i\int_{r_0}^{r'}\left[K_{22}(r'')-K_{11}(r'')dr''\right]}K_{21}(r')\right|^2 \,.
\label{eq:inhomo_prob}
\end{align}    
\end{widetext}

\section{Application to Other Types of Particle Conversions}
\label{sec:otherconversions}

In the previous sections, we presented the unitary evolution method for dark photon and gravitational wave conversions, which does not meet the WKB approximation requirement. Now, we will apply the unitary evolution method to a slightly different but general operator for the EOMs, which does not contain derivatives in the off-diagonal term. This general operator can describe axion-photon conversion under an external magnetic field~\cite{Raffelt:1987im, Dessert:2019sgw} and photon-dark photon conversion in the plasma environment~\cite{Mirizzi:2009iz, Hook:2018iia, An:2020jmf}. Thus, the presented unitary evolution solution is general and has broader applications than the WKB method. Finally, we will show that applying the WKB condition to the unitary evolution solution results in the same solution as the WKB approximation, demonstrating the generality of our unitary evolution solution.

\subsection{Unitary Evolution Solutions for Other Particle Conversions }

We focus on the conversion between two types of fields, namely \( A \) and \( B \). These fields could represent pseudoscalar bosonic particles like axions, vector fields like photons and dark photons, or gravitational wave fields. The masses of particles \( A \) and \( B \) are \( m_A \) and \( m_B \), respectively. Suppose \( A \) is the incoming particle, with energy \( \omega > m_A \).

The EOMs for these fields have the general form:
\begin{equation}
\hat{f}_M \cdot \psi(r) \equiv
\begin{pmatrix}
\Box+m_A^2 & M\\
M^* & \Box+m_B^2
\end{pmatrix} \cdot
\begin{pmatrix}
A\\
B\\
\end{pmatrix}
=0,
\label{eq:AB}
\end{equation}
where the diagonal terms are the free particle dispersion relations for \( A \) and \( B \). \(\hat{f}_M\) is the second-order differential operator for the EOMs, and \(\psi(r)\) is the solution to the two fields. The off-diagonal term \( M \) represents the conversion interactions between \( A \) and \( B \). 

For the photon/dark photon-gravitational wave conversion in Eq.~\eqref{eq:f2x2}, \( M \) contains the spatial derivative. The unitary evolution solution for this case has been given in Eqs.~\eqref{eq:homo_res_K} and \eqref{eq:homo_res}.

Alternatively, we consider a different case where \( M(r) \) can be a function of \( r \) but does not contain its derivative. This is exactly the case for axion-photon conversion and photon-dark photon conversion.

The axion-photon conversion has been extensively studied in Refs.~\cite{OHare:2020wum, Caputo:2020quz, Raffelt:1987im, Hook:2018iia, Dessert:2019sgw, Battye:2019aco, McDonald:2023ohd}. 
The interacting Lagrangian for the axion is given as
\begin{align}
\mathcal L_{a\leftrightarrow \gamma}^{\rm int} = -\frac{1}{4} g_{a\gamma} a \mathbf{E} \cdot \mathbf{B}.
\end{align}
In our formalism, we replace \(A \to \gamma_{||}\) (the photon field), \(B \to a\) (the axion field), and the conversion term $M$ due to axion-photon interaction under the magnetic field,
\begin{align}
    M = g_{a\gamma} \mathbf{B}_{\bot} \omega, \quad \text{for axion-photon conversions.} 
\end{align}

For dark photon-photon oscillation, the interaction Lagrangian in the flavor basis is 
\begin{align}
\mathcal L_{A'\leftrightarrow \gamma}^{\rm flavor} = -\epsilon m_{A'}^2 A'_\mu A^\mu .
\end{align}
Upon replacing \(A \to A'\), \(B \to \gamma\), we have the conversion term $M$ as
\begin{align}
    M = \epsilon m_{A'}^2 \,, \quad \text{for dark photon-photon conversions.} 
\end{align}

By performing the calculation method described in Sec.~\ref{sec:unitary-evolution-solution}, we can find the unitary evolution solution satisfying \(\bar{f}_M \exp{[i \mathbf{K} (r-r_0)]} \bar{\psi}(r_0) = 0\). The Hermitian matrix \(\mathbf{K}\) has the following form
\begin{align}
 \textbf{K}=\begin{pmatrix}
 K_{11}&K_{12}\\
 K_{21}&K_{22}
 \end{pmatrix},
 \end{align} 
with
\begin{align}
K_{11} &= \frac{1}{\sqrt{2}} \left[ \frac{(\chi - m_A^2 + m_B^2)^2 \sqrt{\chi - m_A^2 - m_B^2 + 2\omega^2}}{4|M|^2 + (\chi - m_A^2 + m_B^2)^2} \right. \nonumber \\
&\quad \left. + \frac{(\chi + m_A^2 - m_B^2)^2 \sqrt{-\chi - m_A^2 - m_B^2 + 2\omega^2}}{4|M|^2 + (\chi + m_A^2 - m_B^2)^2} \right], \nonumber \\
K_{22} &= \frac{4|M|^2}{\sqrt{2}} \left[ \frac{\sqrt{2\omega^2 - \chi - m_A^2 - m_B^2}}{4|M|^2 + (\chi + m_A^2 - m_B^2)^2} \right. \nonumber \\
&\quad \left. + \frac{\sqrt{2\omega^2 + \chi - m_A^2 - m_B^2}}{4|M|^2 + (\chi - m_A^2 + m_B^2)^2} \right], \nonumber \\
K_{12} &= K_{21}^* \nonumber \\
&= \frac{M}{\sqrt{2} \chi} \left[ \sqrt{2\omega^2 + \chi - m_A^2 - m_B^2} \right. \nonumber \\
&\quad \left. - \sqrt{2\omega^2 - \chi - m_A^2 - m_B^2} \right],
\label{eq:general_k}
\end{align}

where $\chi \equiv \sqrt{4|M|^2 + (m_A^2 - m_B^2)^2}$.

It is important to note that if \( M(r) \) does not contain the spatial derivative, the unitary evolution solution does not require the rescale factor \(v\). In this scenario, the spatial derivative appears as \(-\partial_r^2 \times \mathbf{I} \to k^2 \times \mathbf{I}\) in the diagonal terms only. Solving the zero determinant condition, \({\rm Det}[f_M(k)]=0\), follows the standard procedure of linear algebra. The two eigenvectors, \(\nu_1\) and \(\nu_2\), are automatically orthogonal, thus eliminating the need to rescale the fields. Consequently, for axion-photon and dark photon-photon conversions, there is no need to redefine the fields, unlike in the case of GW conversion.

The conversion probability can be derived via perturbation theory for the inhomogeneous environment:
\begin{align}
P(A\to B)=\left|\int_{r_0}^rdr'e^{-i\int_{r_0}^{r'}\left[K_{11}(r'')-K_{22}(r'')dr''\right]}K_{21}(r')\right|^2.
\label{eq:inhomo_prob_general}
\end{align}

The above results are general and valid without requiring the WKB approximation condition. We expect these results to reduce to the WKB approximation solutions when the condition for the WKB approximation is satisfied.

\subsection{The WKB approximation}
\label{sec:wkb-approx}

We first review the condition for using the WKB approximation. Assuming radial plane wave solutions, the field can be separated as \( A(r,t) = A(r) e^{ - i \omega t}\equiv \widetilde{A}(r) e^{i (k r - \omega t)} \). The WKB approximation is commonly used to reduce second-order differential equations to first-order differential equations, greatly simplifying the calculation. Under the WKB approximation condition,
\begin{align}
| \partial_r^2 \widetilde{A}(r) | \ll | i k \partial_r \widetilde{A}(r) |,
\label{eq:wkb}
\end{align}
the second order differential can be reduced to first order differential as
\begin{align}
  \partial_r^2 A &= (- k^2 \tilde{A} + 2 i k \partial_r \tilde{A} + \partial_r^2 \tilde{A}) e^{- i (\omega t - k r)} \nonumber \\
  &\approx (- k^2 \tilde{A} + 2 i k \partial_r \tilde{A}) e^{- i (\omega t - k r)}.
  \label{eq:wkb-simplification}
\end{align}

Applying the simplification \eqref{eq:wkb-simplification}, the second-order EOMs can be reduced to a first-order linear differential equation of the general form: 
\begin{equation}
  \left[ - i \partial_z + \mathbf{\Delta} + \mathbf{\Gamma} \right] \psi = 0,
\end{equation}
where \(\psi\) is the field vector, \(\mathbf{\Delta} = \text{diag}(\Delta_1, \ldots, \Delta_n)\) is the diagonal matrix, and \(\mathbf{\Gamma}\) is the off-diagonal matrix, treated as small perturbations. 
As an example, in the two-component case \eqref{eq:AB}, assuming an incoming particle A and applying the WKB approximation, we have:

\begin{align}
\mathbf{\Delta} = \frac{1}{2k}
    \begin{pmatrix}
        0 & 0 \\
        0 & M_B^2 - M_A^2
    \end{pmatrix},
\quad
\mathbf{\Gamma} = \frac{1}{2k}
    \begin{pmatrix}
        0 & M \\
        M^* & 0
    \end{pmatrix}.
\end{align}

For an initial state \((1, \ldots, 0)^{\rm T}\), using perturbation calculation, one can obtain the WKB solution for \(n \neq 1\) as~\cite{Raffelt:1987im}
\begin{equation}
\begin{aligned}
A_n(r) =& - i \int_0^r dr' \, \Gamma_{1n}^{*}(r') \nonumber\\
& \times \exp \left[ i \int_0^{r'} dr'' \, (\Delta_n(r'')  - \Delta_1(r'')) \right].
\end{aligned}
\label{eq:wkb-solution}
\end{equation}

As a consistency check, we substitute the WKB solution \eqref{eq:wkb-solution} back into the WKB condition \eqref{eq:wkb}. The WKB condition requires that
\begin{equation}
  \left| i \left[ \Delta_n(r) - \Delta_1(r) \right] + \frac{d \Gamma_{1 n}^{*} / dr}{\Gamma_{1 n}^{*}} \right| \ll k,
  \label{eq:wkbrefine}
\end{equation}
where the diagonal elements \(\Delta_i\) typically represent plasma frequencies (for photons) or particle masses. The WKB condition can be met for two typical cases:

$\bullet$ \textbf{Case A: Relativistic Particle Conversions}. In this case, we consider particle conversions involving relativistic particles, where the momentum \(k\) is much larger than the characteristic frequencies or masses of the particles involved, i.e., \(k \sim \omega \gg \Delta_i\), and \(\left| d \Gamma_{1 n}^{*} / dr \right| / \Gamma_{1 n}^{*}\) is small, as discussed in Refs.~\cite{Raffelt:1987im, Domcke:2020yzq, Caprini:2019egz, Dessert:2019sgw, Liu:2023mll}. This is commonly the scenario in high-energy astrophysical or cosmological environments.

Using relativistic approximation $\omega\gg m_A$, one can significantly simplify the parameters in Eq.~\eqref{eq:general_k} as 
\begin{align}
K_{11}-K_{22}&\to \frac{m_A^2-m_B^2}{2\omega}+\mathcal{O}\left(\omega^{-2} \right),\nonumber\\
K_{21}&\to  \frac{M}{2\omega}+\mathcal{O}\left(\omega^{-2}\right).
\end{align}
Interestingly, this form coincides with the neutrino oscillation Hamiltonian, which describes the unitary evolution of the neutrino states.

Next, we apply this to a concrete example: the axion-photon conversion under a static external magnetic field. We replace \(A \to a\), \(B \to \gamma_{||}\), and \(M \to g_{a\gamma} \mathbf{B}_{\bot} \omega\) and  arrive at the conversion probability using Eq.~\eqref{eq:inhomo_prob_general}:
\begin{align}
P(\gamma_{||} \to a)^{\rm relativistic} = \left| \int_{r_0}^r dr' \frac{g_{a\gamma} \mathbf{B}_{\bot}}{2} e^{-i \int_{r_0}^{r'} \left[\frac{m_\gamma^2(r'') - m_a^2}{2\omega} dr''\right]} \right|^2 .
\label{eq:gamma2axion_relativistic}
\end{align}
One can check that the above conversion probability exactly matches the results in Refs.~\cite{Dessert:2019sgw, Raffelt:1987im} with the WKB approximation.

$\bullet$ \textbf{Case B: Resonant Particle Conversions.} In resonant conversions, the particle masses are nearly degenerate, \(m_A \approx m_B\). The resonant conversion condition is satisfied when \(\Delta_n(r) \approx \Delta_1(r)\). In the thin resonant layer, the magnetic field or plasma frequency can be considered homogeneous, making the change in \(\Gamma_{1 n}\) mild, i.e., \(d \Gamma_{1 n}^{*} / dr \sim 0\), as discussed in Refs.~\cite{Mirizzi:2009iz, Hook:2018iia, An:2020jmf, Battye:2019aco}.

In this case, we have the following simplifications based on Eq.\eqref{eq:general_k} :
\begin{align}
K_{11} - K_{22} &\to \frac{(m_B - m_A) m_A}{\sqrt{\omega^2 - m_A^2}} + \mathcal{O}(m_B - m_A)^2\nonumber\\
&\quad \approx \frac{m_B^2 - m_A^2}{2k_A} + \mathcal{O}(m_B - m_A)^2, \nonumber \\
K_{21} &\to \frac{M}{2k_A},
\end{align}
where we define $k_A = \sqrt{\omega^2 - m_A^2}$ which is the on-shell momentum of the incoming particle. 
For illustration, we choose axion-photon and dark photon-photon conversions.

For non-relativistic axion dark matter, we have \(\omega \sim m_a\). The axion-photon conversion in the plasma environment dominantly happens when \(m_a \sim \omega_{\rm pl}\). We replace \(A \to a\), \(B \to \gamma_{||}\), and \(M \to g_{a\gamma} \mathbf{B}_{\bot} \omega\). Then, the conversion probability is:
\begin{align}
P(a \to \gamma_{||})^{\rm resonant} = \left| \int_{r_0}^r dr' \frac{g_{a\gamma} \mathbf{B}_{\bot} \omega}{2k_a} e^{-i \int_{r_0}^{r'} \left[\frac{m_a^2 - m_\gamma^2(r'')}{2k_a} dr''\right]} \right|^2 \,.
\label{eq:gamma2axion_resonant}
\end{align} 
Our results exactly reduce to the results in Ref.~\cite{Hook:2018iia} using the WKB approximation.

For dark photon-photon oscillation, the interaction Lagrangian in the flavor basis is 
\begin{align}
\mathcal L_{A'\leftrightarrow \gamma}^{\rm flavor} = -\epsilon m_{A'}^2 A'_\mu A^\mu .
\end{align}
We consider dark photon dark matter converted into photons in a plasma environment. The plasma mass varies with \(r\) due to the electron number density distribution. The conversion is dominated by the resonant conversion in a thin layer when \(m_{A'} \approx \omega_{\rm pl}\)~\cite{An:2020jmf}. Upon replacing \(A \to A'\), \(B \to \gamma\), and \(M \to \epsilon m_{A'}^2\), we obtain the conversion probability as
\begin{align}
P(A' \to \gamma)^{\rm resonant} = \left| \int_{r_0}^r dr' \frac{-\epsilon m_{A'}^2}{2k} e^{-i \int_{r_0}^{r'} \left[\frac{m_{A'}^2 - m_\gamma^2(r'')}{2k} dr''\right]} \right|^2 \,.
\label{eq:dp2p_resonant}
\end{align}
This matches the results in Refs.~\cite{Hook:2018iia, An:2020jmf} using the WKB approximation.

In summary, we have explicitly shown that the general unitary evolution solution can reduce to the WKB solutions when the WKB condition is applied, as illustrated in Cases A and B.

\section{Characteristic Strain Calculation}
\label{sec:strain}

In this section, we calculate the strength of gravitational waves converted from dark photon dark matter within the solar magnetic field. We will work in the flavor basis, using the conversion probability in Eq.~\eqref{eq:inhomo_prob}. Since the photon degree of freedom has been integrated out, we only need to consider the external sterile magnetic field $\mathbf{B}_{\rm S}$. The strength of the background sterile magnetic field $\mathbf{B}_{\rm S}$ can be calculated through the EOMs involving photons and dark photons, with the active magnetic field profile of the Sun $\mathbf{B}_{\rm I}(r)$. For the leading-order calculation, we consider flat space. Once we obtain $\mathbf{B}_{\rm S}$, we will calculate the conversion probability to gravitational waves and estimate the strength of the characteristic strain for gravitational wave experiments.

\subsection{The Sterile Magnetic Field $\bs$ Inside the Sun}
For the photon and dark photon (Eq.~\eqref{eq:dp}), assuming a static interactive field $A_I$, we can obtain the EOM for the sterile magnetic field $\mathbf{B}_{\rm S}^i = \frac{1}{2} \epsilon^{ijk} F_{\rm S}^{jk}$ as
\begin{equation}
\vec{\nabla} \times \mathbf{B}_{\rm S} = -m_{A'}^2 \mathbf{A}_{\rm S} + \epsilon m_{A'}^2 \mathbf{A}_{\rm I},
\end{equation}
where the right-hand side term is an effective current from the mass term and the mass mixing term. This indicates that in the flavor basis, a non-zero active magnetic field $\mathbf{A}_{\rm I}$ will induce a dark magnetic field $\mathbf{B}_{\rm S}$. Taking the curl on both sides of the equation leads to,
\begin{equation}
    \vec{\nabla}^2 \mathbf{B}_{\rm S}-m_{A'}^2 \mathbf{B}_{\rm S}=-\epsilon m_{A'}^2 \mathbf{B}_{\rm I},
    \label{eq:bp}
\end{equation}
where we have used $\vec{\nabla}\times \left(\vec{\nabla}\times\mathbf{B}_{\rm S}\right)= \vec{\nabla} \left(\vec{\nabla} \cdot \mathbf{B}_{\rm S}\right) -\vec{\nabla}^2 \mathbf{B}_{\rm S}$, $\vec{\nabla} \cdot \mathbf{B}_{\rm S/I} = 0$, $\mathbf{B}_{\rm I} = \vec{\nabla} \times \mathbf{A}_{\rm I}$, $\mathbf{B}_{\rm S} = \vec{\nabla} \times \mathbf{A}_{\rm S}$. Equipped with the profile of the interactive magnetic field $\mathbf{B}_{\rm I}$, one can directly use the Green's function method to calculate its value at a certain point $\mathbf{r}$:
\begin{equation}
    \mathbf{B}_{\rm S}(\mathbf{r})=\frac{\epsilon m_{A'}^2}{4\pi} \int d^3 \mathbf{r'}\frac{\mathbf{B}_{\rm I}(\mathbf{r'})}{|\mathbf{r}-\mathbf{r'}|} e^{-m_{A'}|\mathbf{r}-\mathbf{r'}|} .
    \label{eq:gr}
\end{equation}

\begin{figure}
\centering
\includegraphics[width= 0.99 \linewidth]{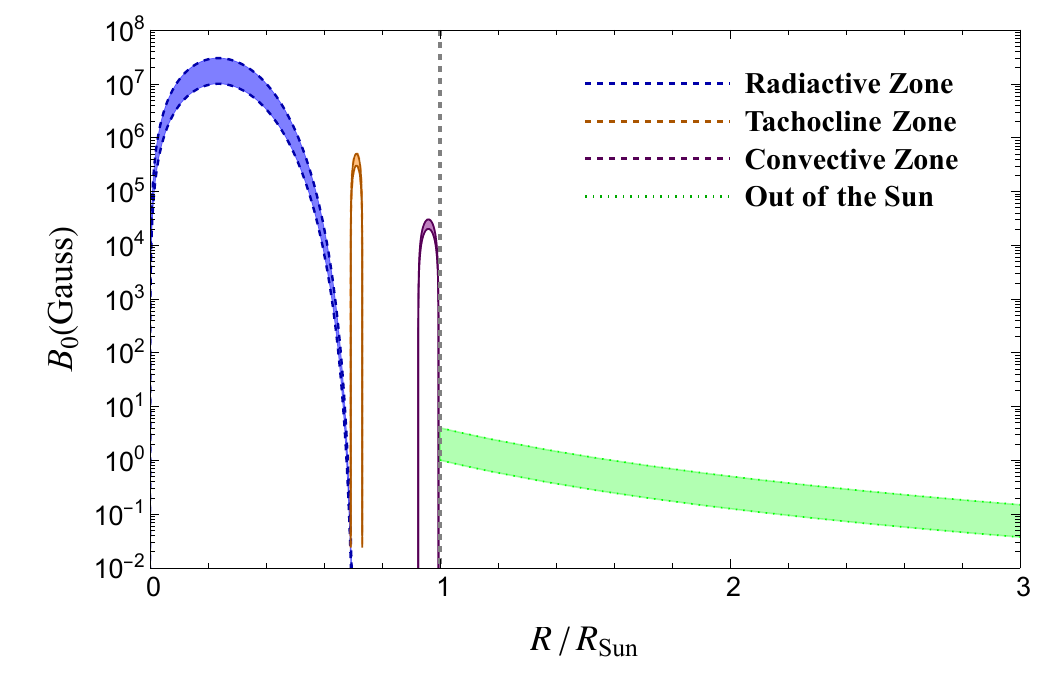}
\caption{Profile of the solar magnetic field $B_0(r)$ as a function of the normalized radius $r/R_{\rm Sun}$ with parameter uncertainties demonstrated as finite bandwidth.  The explicit profile for each part of the field within the Sun is given by Eq.~\eqref{eq:sunb}, which could be found in Refs.~\cite{OHare:2020wum, Caputo:2020quz, Antia:2000pu,Couvidat:2002gvk,Guarini:2020hps}. Outside the solar radius, the magnetic field follows a dipole-like profile with fluctuations~\cite{ASCHWANDEN2014235,Solanki:2006cyh}. To proceed conservatively, we adopt the magnetic field strength to be $1 \sim 4$ Gauss at the radius of the Sun.}
\label{fig:sun_mag} 
\end{figure} 

In our study, $\mathbf{B}_{\rm I}(r)$ represents the standard active magnetic field of the Sun from Refs.~\cite{OHare:2020wum, Caputo:2020quz, Antia:2000pu, Couvidat:2002gvk, Guarini:2020hps}, the magnitude of which is shown in Fig.~\ref{fig:sun_mag}. To simplify, we assume spherical symmetry for the solar magnetic field, which is determined by a radial profile $B_0(r)$ given by $\mathbf{B}_{\rm I}(\mathbf{r}) = B_0(r) \sin{\theta} \hat{\phi}$ with $\hat{\phi}$ being the unit vector in the $\phi$ direction (azimuthal angle). The function $B_0(r)$ is characterized by three distinct regions: the radiative zone, tachocline, and convective zone. The function $B_0(r)$ is described by the following equations:
\begin{widetext}
\begin{align}
B_0(r)=\begin{cases}
  &\displaystyle B_{\rm rad} K_\lambda\left(\frac{r}{r_1}\right)^2\left[1-\left(\frac{r}{r_1}\right)^2\right]^\lambda ,~~~0\ll r\le r_1~~~\text{radiactive zone}\\
  &\displaystyle B_{m,1}\left[1-\left(\frac{r-r_1}{d_1}\right)^2\right],~~~|r-r_1|\le d_1~~~\text{tachocline zone}\\
  &\displaystyle B_{m,2}\left[1-\left(\frac{r-r_2}{d_2}\right)^2\right],~~~|r-r_2|\le d_2~~~\text{convective zone}\\
\end{cases}
\label{eq:sunb}
\end{align}    
\end{widetext}
where $ r_1 = 0.712 R_\odot$, $r_2 = 0.96 R_\odot$, $d_1 = 0.02 R_\odot$, and $d_2 = 0.035 R_\odot$ with the solar radius $R_\odot = 6.9598 \times 10^8 \text{ m}$ \cite{Serenelli:2009yc}. We have $10^7 \text{ G} \lesssim B_{\rm rad} \lesssim 3 \times 10^7 \text{ G}$, $K_\lambda = (1+\lambda)(1+1/\lambda)^\lambda$, with $\lambda \equiv 1 + 10r_1/R_\odot$ for the radiative zone \cite{Couvidat:2002gvk}, $3 \times 10^5 \text{ G} \lesssim B_{m,1} \lesssim 5 \times 10^5 \text{ G}$ for the tachocline zone, and $2 \times 10^4 \text{ G} \lesssim B_{m,2} \lesssim 3 \times 10^4 \text{ G}$ for the convective zone \cite{Antia:2000pu}.

To efficiently estimate the $\mathbf{B}_{\rm S}$ field without sacrificing precision, we introduce the following simplification methods by identifying three distinct mass regions characterized by the relationship between $m_{A'}$ and $1/R_\odot$:

\begin{itemize}
\item[$\bullet$] For $m_{A'} \leq 1/R_\odot$, or equivalently $f_{A'} \leq 0.1~\text{Hz}$ (Simplified Method 1), the exponential term $e^{-m_{A'}|\mathbf{r}-\mathbf{r'}|}$ in the integrand of Eq.~\eqref{eq:bp} is negligible, allowing us to consider a slowly varying configuration of $\mathbf{B}_{\rm S}$. In this scenario, numerical solutions to the differential equations can be utilized. Given the spherical symmetry of the solar magnetic field $\mathbf{B}_{\rm I}$, we can hypothesize that $\mathbf{B}_{\rm S}$, which satisfies the field equations, also exhibits spherical symmetry $\mathbf{B}_{\rm S} = b(r) \sin{\theta} \hat{\phi}$. The justification of the ansatz is presented in the Appendix~\ref{sec:Bs}. This reduces the field equations to a single radial differential equation for the profile $b(r)$:

\begin{equation}
    \frac{d^2 b(r)}{dr^2}+\frac{2}{r}\frac{d b(r)}{dr^2}- \left(\frac{2}{r^2}+m_{A'}^2 \right)b(r)=-\epsilon m_{A'}^2B_0(r),
    \label{eq:reduced-r-eq}
\end{equation}
with the boundary condition provided by evaluation of Green function Eq.~\eqref{eq:gr} at specific points.

\item[$\bullet$] For $f_{A'} \in [0.1, 10]~\text{Hz}$ (Simplified Method 2), the exponential part of the integrand in Eq.~\eqref{eq:gr} begins to make a noticeable contribution but is still not dominant. In this range, one can efficiently determine $\mathbf{B}_{\rm S}$ by utilizing the Green's function in Eq.~\eqref{eq:gr} and interpolating the results.

\item[$\bullet$] For $f_{A'} \geq 10~\text{Hz}$ (Simplified Method 3), corresponding to the large mass limit $m_{A'} \gg 1/R_\odot$, we can approximate $\mathbf{B}_{\rm I}(\mathbf{r}')$ around $\mathbf{r}$ by defining $\mathbf{a} \equiv \mathbf{r}' - \mathbf{r}$ and expanding as follows:
\begin{equation}
    \bi(\mathbf{r}')=\bi(\mathbf{r})+\mathbf{a}\cdot \nabla_{\mathbf{r}} \bi(\mathbf{r})+\frac{1}{2}(\mathbf{a}\cdot \nabla_{\mathbf{r}})^2 \bi(\mathbf{r})+\dots.\\
    \label{eq:expand}
\end{equation}
Since the function is exponentially suppressed at large $\mathbf{a}$, we can retain only the leading terms in the expansion. Substituting this expansion into Eq.~\eqref{eq:gr}, we can solve the fields order by order as $\mathbf{B}_{\rm S}(\mathbf{r}) \approx \mathbf{B}_{0}'(\mathbf{r}) + \delta \mathbf{B}'(\mathbf{r})$. The leading order term $\mathbf{B}_{0}'(\mathbf{r})$ is:
\begin{equation}
    \mathbf{B}'_0(\mathbf{r})=\frac{\epsilon m_{A'}^2}{4\pi} \int d^3 \mathbf{a}\frac{\bi(\mathbf{r})}{|\mathbf{a}|} e^{-m_{A'}|\mathbf{a}|}=\epsilon \bi(\mathbf{r}).\\
\end{equation}
The second term in Eq.~\eqref{eq:expand} leads to zero results because the integrand is odd in $\mathbf{a}$. Therefore, the next leading-order contribution comes from the third term:
\begin{equation}
    \delta \mathbf{B}'(\mathbf{r})=\frac{\epsilon m_{A'}^2}{4\pi} \int d^3 \mathbf{a}\frac{1}{2|\mathbf{a}|} \left[ \left( \mathbf{a}\cdot \nabla_{\mathbf{r}} \right)^2 \mathbf{B}_{\rm I}(\mathbf{r})\right]e^{-m_{A'}|\mathbf{a}|}.
\end{equation}

Substituting the specific magnetic field profile $\mathbf{B}_{\rm I}(\mathbf{r}) = B_0(r) \sin{\theta} \hat{\phi}$, we obtain:

\begin{equation}
    \delta \mathbf{B}'(\mathbf{r})=\frac{\epsilon}{m_{A'}^2} \left(-\frac{B_0(r)}{3r^2}+\frac{2}{3r}\frac{dB_0(r)}{dr}+\frac{1}{3}\frac{d^2B_0(r)}{dr^2} \right) \hat{\phi}.
\end{equation}
This result indicates that for large $m_{A'}$ (compared to the inverse of the magnetic field range), the dark magnetic field $\mathbf{B}_{\rm S}$ generated by the ordinary magnetic field $\mathbf{B}_{\rm I}$ differs from $\mathbf{B}_{\rm I}$ by a factor of $\epsilon$, which is consistent with physical intuition.

\end{itemize}

Outside the solar radius, the solar magnetic field reduces to a tiny value of around 1 Gauss. Numerical solutions of the EOMs~\eqref{eq:dp} confirm that the conversion probability outside the Sun is negligible compared to that inside the Sun. Thus, for simplicity, we consider only the conversion process within the Sun.

\begin{figure*}
\centering
\includegraphics[width= 0.49 \linewidth]{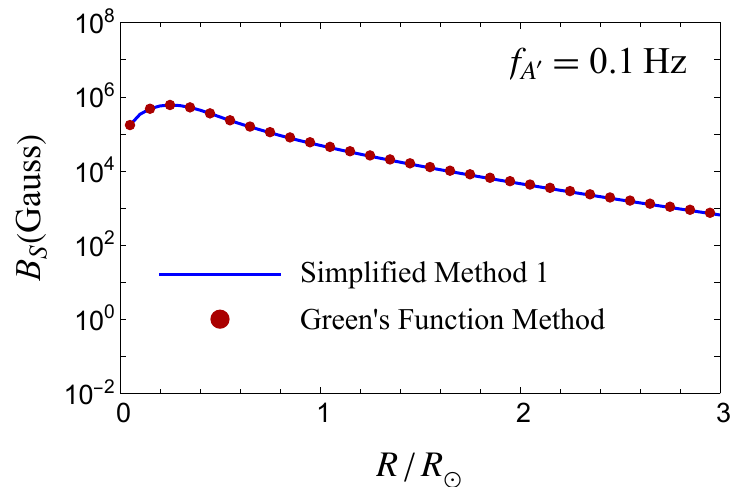}
\includegraphics[width= 0.49 \linewidth]{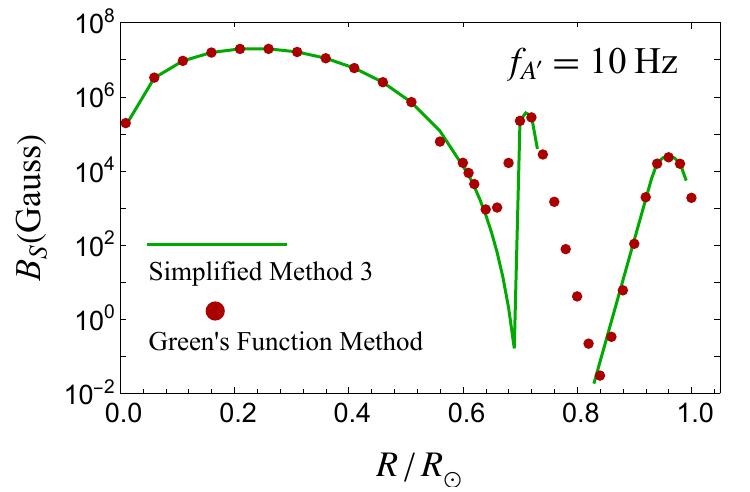}
\caption{
The calculated sterile magnetic field profiles using the Green's function method and the simplified methods are shown for $f_{A'} = 0.1$ and $10$ Hz, respectively, with $\epsilon=1$.
}
\label{fig:compare_res} 
\end{figure*}

In Fig.~\ref{fig:compare_res}, we compare the calculated sterile magnetic field $\mathbf{B}_{\rm S}$ from numerically integrating the Green's function and the simplified methods 1 and 3 for $f_{A'} = 0.1$ and $10$ Hz, respectively, with \(\epsilon=1\). While the Green's function is believed to provide an accurate evaluation of $\mathbf{B}_{\rm S}$, it requires significant computational resources to compute the entire $\mathbf{B}_{\rm S}$ profile. To efficiently estimate the $\mathbf{B}_{\rm S}$ field without sacrificing precision, we have introduced two additional methods to simplify the calculation. By comparing our simplified methods with the Green's function results in the corresponding mass range, we demonstrate the effectiveness of our approximation methods. The left panel illustrates the results for small $f_{A'}$ ($= 0.1~\text{Hz}$) compared with the Green's function result (red dot) and the Simplified Method 1 (blue line) numerical solution to the differential equation. The right panel illustrates the results for large $f_{A'}$ ($= 10~\text{Hz}$) compared with the Green's function result (red dot) and the Simplified Method 2 (green line) using the power expansion.

\subsection{Calculations of GW Characteristic Strain}

The dark photon dark matter field corresponds to $A'$ in the mass basis, because it exists for a very long time and the coherent flavor states should already decohere into mass states. When transformed into the flavor basis, the initial state amplitude is thus given by  
$\bar{\psi}(r_0)=\sqrt{v} A_{\rm DM}\{-\epsilon,1,0\}^{\rm T}$, where 
$A_{\rm DM} = (2 \rho_{\rm DM})^{1/2} / m_{A'}$ represents the amplitude of incoming dark photons with $\rho_{\rm DM}$ being the energy density of DPDM.

We can calculate the GW production through Eq. (\ref{eq:inhomo_prob}) as follows:
\begin{align}
   h^2 &= \kappa^2 \tilde{h}^2 = \kappa^2\left|\sqrt{v} A_{\rm DM}\right|^2 P(A' \rightarrow {\rm GW}) \nonumber \\
   &= \kappa^2 v \frac{2\rho_{\rm DM}}{m_{A'}^2} \left|\int_{r_0}^r dr' e^{-i\int_{r_0}^{r'} \left[K_{22}(r'') - K_{11}(r'') dr''\right]} K_{21}(r')\right|^2,
\end{align}
To estimate the final GW strength detected by a near-Earth detector, we consider a simple case where the direction of the detector is perpendicular to the solar axis. For the dark matter velocity distribution, we take the Standard Halo Model~\cite{Lee:2013xxa} with a Maxwell-Boltzmann velocity distribution $f(\mathbf{v}) d^3\mathbf{v}$. The resulting characteristic strain is given by:
\begin{align}
    h_c^2 &= \left\langle h^2 \right \rangle= \int_{\Delta\Omega} d^3\mathbf{v} f(\mathbf{v})\, \kappa^2 v \frac{2\rho_{\rm DM}}{m_{A'}^2} P(\mathbf{v})\\
    &= \int dv \, \kappa^2 v^3 \frac{2\rho_{\rm DM}}{m_{A'}^2} \langle P(v) \rangle_{\text{angle}} \Delta \Omega,
\end{align}

with 
\begin{align}
   f(\mathbf{v})=(\frac{1}{2\pi v_0^2})^{3/2} e^{-\frac{(\mathbf{v}-\mathbf{v_{A'}})^2}{2 v_0^2}}
\end{align}
where $\mathbf{v_{A'}}$ denotes the average velocity of DPDM relative to the Solar System, with $|\mathbf{v_{A'}}|\approx 10^{-3}$. $\Delta \Omega$ represents the solid angle from the Sun to Earth, and $P(\mathbf{v})$ is the conversion probability of a dark photon particle traveling through the Sun with velocity $\mathbf{v}$.
The solid angle-averaged $\langle P(v) \rangle_{\text{angle}}$ is defined as
\begin{align}
    \langle P(v) \rangle_{\text{angle}} \equiv \int d\Omega \, f(\mathbf{v}) \, \frac{P(\mathbf{v})}{\Delta \Omega}.
\end{align}
For a rough estimation, we simply take $\langle P(v) \rangle_{\text{angle}} \approx \frac{1}{2} P_{0}(v)$, where $P_{0}(v)$ is the conversion probability of DPDM traveling along the equator of the Sun. 

\begin{figure}
\centering
\includegraphics[width= 0.99 \linewidth]{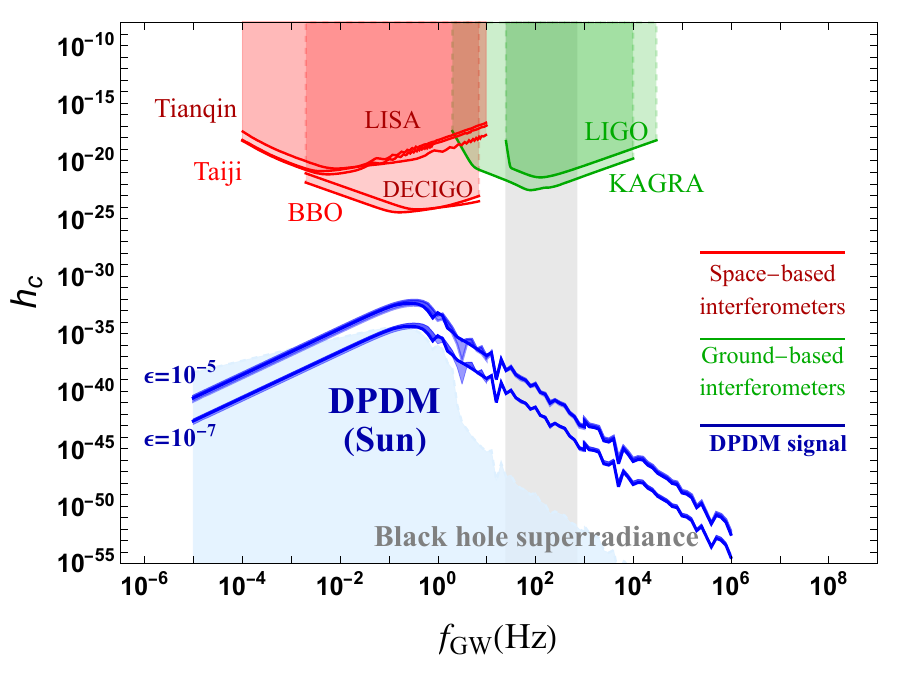}
\caption{The projected characteristic strains $h_c$ of dark photon dark matter converted gravitational waves and associated uncertainties are represented as a blue band with $\epsilon$ values of $10^{-5}$ and $10^{-7}$ are shown, with the parameter region of  characteristic strains compatible with allowed $\epsilon$ are shown as light blue shaded region. The sensitivity curves of existing gravitational wave detectors are shown, with space-based interferometers (red) including TianQin~\cite{TianQin:2015yph}, LISA~\cite{LISA:2017pwj}, DECIGO/BBO~\cite{Yagi:2011wg}, Taiji~\cite{10.1093/ptep/ptaa083}, we adopt the latest sensitivity results from ref.~\cite{Yu:2023iog} ,and ground-based interferometers (green) including LIGO~\cite{LIGOScientific:2014pky} and KAGRA~\cite{Somiya:2011np}.
The black hole superradiance constraint~\cite{Cardoso:2018tly} is shown as a gray shaded region.}
\label{fig:hc_res} 
\end{figure}

In Fig.~\ref{fig:hc_res}, we present the gravitational wave field strength $h_c$ converted from the dark photon dark matter through the large magnetic field inside the Sun. We choose the kinetic mixing parameter between photon and dark photon to be $\epsilon = 10^{-5}, 10^{-7}$ for demonstration. The Compton frequency of the dark photon ranges from $10^{-5}~\text{Hz}$ to $10^{3}~\text{Hz}$, peaking at approximately $0.3~\text{Hz}$ when the Compton wavelength is comparable to the Sun's radius. We also select the maximum value of $\epsilon$ as a function of $m_{A'}$, permitted by all current constraints on dark photon dark matter (see Ref.~\cite{Caputo:2021eaa} and the \href{https://cajohare.github.io/AxionLimits/docs/dp.html}{AxionLimits} repository). The corresponding $h_c$ is shown in the light blue shaded region. Finally, we present the sensitivity of various GW interferometers. For the illustrated frequency range, it appears unlikely that dark photon dark matter will be detected by GW interferometer experiments in the near future. Nonetheless, we offer an exotic GW source from DPDM conversion via inner solar magnetic fields.

\subsection{Comparison with the WKB solution}
\label{sec:result}

In the case of non-relativistic dark photon dark matter converting into gravitational waves within a solar magnetic field, the condition for the WKB approximation (Eq.~\eqref{eq:wkbrefine}) does \textit{not} hold. However, the WKB approximation is a commonly used method for particle conversions or oscillations. Therefore, we compare our unitary evolution method results with the WKB approximation results to evaluate the extent of the difference between them.

Using the WKB approximation, the EOMs can be simplified into first-order differential equations in the flavor basis as 
\begin{equation}
    \left[-i\partial_{l}+\frac{1}{2k}
    \begin{pmatrix}
        m_{A'}^2-\Delta_{\rm I} & \epsilon m_{A'}^2 & i \kappa B_{\rm I}k\\
        \epsilon m_{A'}^2 & 0 & i \kappa B_{\rm S}k\\
        -i\kappa B_{\rm I}k & -i\kappa B_{\rm S}k & m_{A'}^2
    \end{pmatrix}
    \right]\begin{pmatrix}
        A_{\rm I}^\lambda\\A_{\rm S}^\lambda\\ \tilde{h}^\lambda
    \end{pmatrix}=0.
\end{equation}
Since $\Delta_{\rm I} \gg m_{A'}$ and perturbativity holds, the leading-order conversion probability for $A' \to {\rm GW}$ is given by
\begin{align}
P_{\rm WKB}( A' \to {\rm GW}) = \left|\int_{r_0}^r dr' \frac{-i\kappa B_{\rm S}}{2} e^{-i\int_{r_0}^{r'} \left[\frac{m_{A'}^2}{2k} dr''\right]}\right|^2.
\label{eq:wrong}
\end{align}

We can compare this to the full unitary evolution method results in the first line of Eq.~\eqref{eq:homo_res_K} and Eq.~\eqref{eq:inhomo_prob}, where 
\begin{align}
& K_{33} - K_{22} = \frac{(1 - v) \sqrt{B_{\rm S}^2 \kappa^2 + (v + 1)^2 \omega^2}}{v + 1}, \\
& K_{32} = \frac{i \kappa B_{\rm S} \sqrt{v}}{v + 1}.
\end{align}
When taking the relativistic limit $\displaystyle v = \sqrt{1 - \frac{m_{A'}^2}{\omega^2}} \to 1$ or $\omega/m_{A'} \to \infty$, our result can align with the WKB result above in Eq.~\eqref{eq:wrong}. 

Unfortunately, since the DPDM is non-relativistic, we have $v = v_{A'} \ll 1$ , which does not meet the WKB condition. The full unitary evolution method gives the conversion probability as 

\begin{equation}
    P_{\rm{Full}}(A' \to \mathrm{GW}) \approx \left| \int_{r_0}^r dr' (-i\kappa B_{\rm S} \sqrt{v} e^{-i\int_{r_0}^{r'} m_{A'} dr''}) \right|^2,
\end{equation}
where we have assumed $B_{\rm S} \kappa \ll m_{A'}$, indicating a weak dark magnetic field.

With the conversion probability, we can estimate the characteristic strain without averaging the DM velocity profile. To understand the order of magnitude of the strain, we assume the dark photon passes through the Sun's equatorial plane and simplify the solar dark magnetic field as a 1D function \( B_{\rm S}(r) \) with an anti-symmetric profile: \( B_{\rm S}(r) = \text{sign}(r) B_{\rm S}^0 \) within the interval \( r \in [-\Delta r, \Delta r] \), where \( r = 0 \) represents the center of the Sun. Additionally, we assume \( B_{\rm S}^0 \) and \( m_{A'} \) to be constant. Note that the anti-symmetric profile, where \( B_{\rm S}(r) \) changes sign, is a key feature of the realistic environment.

Therefore, the strains can be simplified to:
\begin{align}
   h_{c, \text{WKB}}^2  &= \kappa^2 P_{\rm WKB} \times A_{{\rm S},0}^2 \nonumber\\
   & = \left( \frac{4\kappa^2 B_{\rm S}^0 v}{m_{A'}} \right)^2 \sin^4 \left( \frac{m_{A'} \Delta r}{4v} \right) A_{{\rm S},0}^2 \,,
\label{eq:wkb-simp} \\
h_{c, \text{Full}}^2  &= \kappa^2 P_{\rm Full} \times (\sqrt{v} A_{{\rm S},0})^2 \nonumber\\
& = \left( \frac{4\kappa^2 B_{\rm S}^0 v}{m_{A'}} \right)^2 \sin^4 \left( \frac{m_{A'} \Delta r}{2} \right) A_{{\rm S}, 0}^2 \,,
\label{eq:full-simp}
\end{align}
where $A_{{\rm S}, 0}$ is the initial DPDM field. In the second line, the rescaling factor $\sqrt{v}$ is applied.

\begin{figure*}
\centering
\includegraphics[width= 0.49 \linewidth]{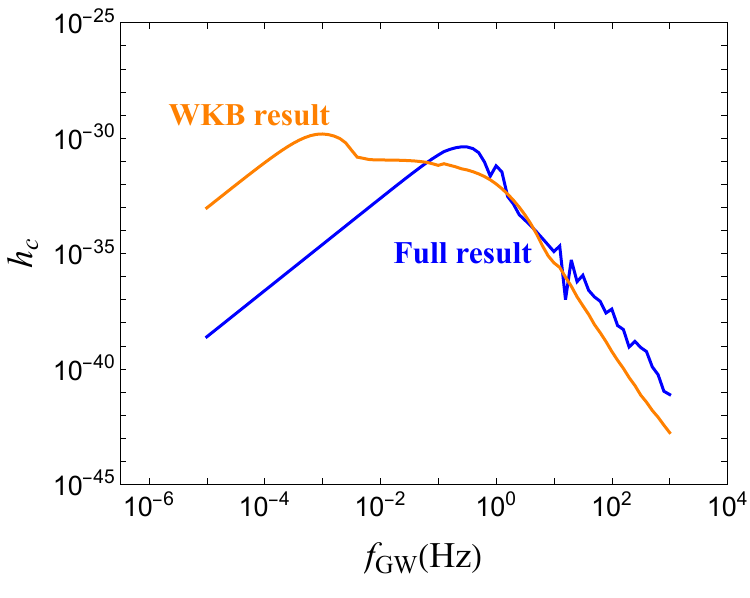}
\includegraphics[width= 0.49\linewidth]{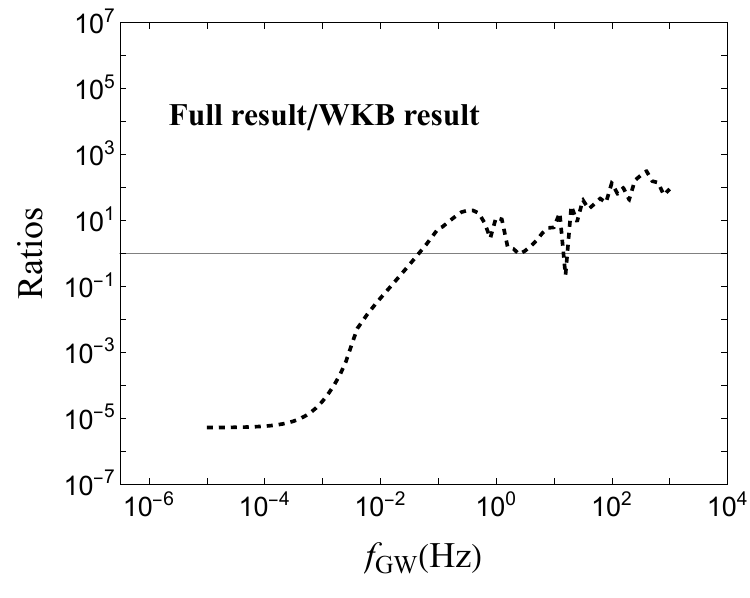}
\caption{The characteristic strains of gravitational waves from the WKB result (orange) and the full unitary evolution result (blue) for \(\epsilon = 10^{-3}\). The right panel shows the ratio of the characteristic strains. 
}
\label{fig:wkb} 
\end{figure*}

Comparing the characteristic strains \( h_{c, \rm WKB} \) and \( h_{c, \rm Full} \) for the simplified case, we observe that the prefactors are identical, but the arguments of the sine function differ. Taking $v=v_{A'}$, the average velocity of DPDM relative to the Sun, the ``oscillation length" is \( 4v_{A'}/m_{A'} \) in the WKB case and \( 2/m_{A'} \) in the full solution case. Consequently, for very small \( m_{A'} \), we have \( h_{c, \rm WKB}^2 / h_{c, \rm Full}^2 \approx v_{A'}^{-4} \). For larger \( m_{A'} \), the WKB case will exhibit a much larger evolution phase in the exponential term of Eq.~\eqref{eq:wrong}, leading to greater amplitude cancellation in the WKB case compared to the Full case. 

In Fig.~\ref{fig:wkb}, we use the realistic solar \( B_{\rm S}(r) \) profile to numerically calculate the conversion probabilities \( P_{\rm{WKB}}(A' \to \mathrm{GW}) \) and \( P_{\rm{Full}}(A' \to \mathrm{GW}) \), and obtain the characteristic strains. As expected from the simplified cases, for small \( m_{A'} \), the ratio of \( h_c \) is approximately \( v_{A'}^{-2} \). For larger \( m_{A'} \), the full evolution result is larger than the WKB result.

We highlight that for DPDM \(\rightarrow\) GW conversion inside the Sun, the WKB approximation condition is \textit{not} satisfied. Using the WKB solution in this context would lead to errors by orders of magnitude compared to the correct full evolution solution.

\section{Conclusions}
\label{sec:conclusion}

In this work, we estimate the generation of gravitational waves by dark photon dark matter traveling through the magnetic field inside the Sun. Since the dark matter is non-relativistic, the WKB approximation is not applicable. We carefully solve the second-order differential equations of motion and found a unitary evolution solution, which is similar to the results in Ref.~\cite{Domcke:2020yzq}. Since the method is general, we also provide unitary evolution solutions for axion-photon and dark photon-photon conversions when the WKB condition is not satisfied. As a consistency check, we work out the WKB condition that holds for two typical cases: the relativistic particle case and the resonant conversion case. Applying the condition of these two cases, we confirm that the unitary evolution solution reduces back to the WKB approximation solution.

Additionally, we numerically calculate the gravitational wave characteristic strain from the dark photon dark matter conversion through the magnetic field inside the Sun. Considering a Compton frequency range of $10^{-5}$ to $10^6$ Hz for the dark photon dark matter, the photon component can be decoupled due to the large plasma mass inside the Sun. For two states, we provide an analytic solution to the conversion probability from the dark photon dark matter to gravitational waves. We present simplified methods tailored to both large and small dark photon masses to calculate the characteristic strain. The resulting strain signal is far below the sensitivity of current gravitational wave interferometers. Nevertheless, we propose an exotic gravitational wave source, especially at high frequencies, from the dark photon dark matter conversion inside the Sun. Even though current constraints on kinetic mixing ($\epsilon$) limit the detectability of dark photon dark matter, theoretical models suggest the possible existence of dark fermions that are charged under the dark photon. This introduces the possibility of dark magnetic fields, which could facilitate the $A' \to \text{GW}$ conversion without suppression from $\epsilon$, thereby greatly enhancing the potential for such a conversion. As a result, although our study may not be feasible for detection under current conditions, it holds significant promise for future experimental relevance, particularly if the dark sector conditions change or if new astrophysical scenarios arise.

\section{Acknowledgement}
The work of J.L. is supported by the National Science Foundation of China under Grants No. 12075005 and No. 12235001, and by Peking University under startup Grant No. 7101502458. The work of X.P.W. is supported by the National Science Foundation of China under Grant No. 12375095, No. 12005009 and the Fundamental Research Funds for the Central Universities.

\appendix

\section{The Derivation of EOMs from Mass Basis}
\label{sec:der-EOM-massbasis}

It is obvious that physical phenomena shall be independent of the choice of calculation basis, \textit{i.e.}, we expect the same physical result presented in main-text at another basis, such as the mass basis~\eqref{eq:massbasis}.

To clearly see the derivation in the mass basis, we begin our calculation with all $\epsilon$-dependent terms restored. The detailed derivation is given below.

The mass basis Lagrangian with full $\epsilon$-dependent term restored is 
\begin{align}
    \mathcal{L}_{{\rm vac},\epsilon}&=-\frac{1}{4}F_{\mu\nu}F^{\mu\nu}-\frac{1}{4}F'_{\mu\nu}F'^{\mu\nu}+\frac{1}{2}m_A'^2A'_\mu A'^\mu \nonumber\\
    &\quad +\frac{1}{\sqrt{1+\epsilon^2}}J_{\rm EM}^{\mu}(A_\mu-\epsilon A'_\mu),
   \label{eq:massbasis_epsilon}
\end{align}

in which we adopt a similar notation from Ref.~\cite{Dubovsky:2015cca} with the unconventional factor $1/\sqrt{1+\epsilon^2}$. Clearly this Eq.~\eqref{eq:massbasis_epsilon} returns to Eq.\eqref{eq:massbasis} for $\epsilon\ll 1$.  The mass basis could be related to flavor basis by a $SO(2)$ rotation:
\begin{align}
\begin{pmatrix}
A\\
A'
\end{pmatrix}=\mathbf{R}
\begin{pmatrix}
A_{\rm I}\\
A_{\rm S}
\end{pmatrix},   ~~~~~\mathbf{R}\equiv\frac{1}{\sqrt{1+\epsilon^2}}
\begin{pmatrix}
1 & ~\epsilon\\
-\epsilon & ~1
\end{pmatrix}
\label{eq:basistrans_full}
\end{align}
\begin{align}
   \mathcal{L}_{{\rm flavor},\epsilon}&=-\frac{1}{4}(F_{\rm I})_{\mu\nu}(F_{\rm I})^{\mu\nu}
   -\frac{1}{4}(F_{\rm S})_{\mu\nu}(F_{\rm S})^{\mu\nu}\nonumber\\
   &\quad +\frac{m_A'^2}{2(1+\epsilon^2)}(A_{\rm S})_\mu(A_{\rm S})^\mu
   -\frac{\epsilon}{\epsilon^2+1} m_A'^2(A_{\rm S})_\mu(A_{\rm I})^\mu\nonumber\\
   &\quad+\frac{\epsilon^2m_{A'}^2}{2(1+\epsilon^2)}A_{\rm I}^2+(A_{\rm I})_\mu J_{\rm EM}^{\mu}.
   \label{eq:massbasis_epsilon}
\end{align}
When entering the plasma medium, the interaction state $A_{\rm I}$ acquires an extra plasma mass term coming from the interaction $(A_{\rm I})_\mu J^\mu_{\rm EM}$ between plasma and $A_{\rm I}^{\mu}$: $\displaystyle\frac{\epsilon^2m_{A'}^2}{2(1+\epsilon^2)}A_{\rm I}^2\to \frac{1}{2}\left(\frac{\epsilon^2m_{A'}^2}{(1+\epsilon^2)}+\omega_{\rm pl}^2\right)A_{\rm I}^2$.  After transforming back to mass basis we have:
\begin{align}
   &\mathcal{L}_{{\rm vac},\epsilon}^{\rm plasma}=\nonumber\\
 &  -\frac{1}{4}F_{\mu\nu}F^{\mu\nu}-\frac{1}{4}F'_{\mu\nu}F'^{\mu\nu}-\frac{\epsilon}{1+\epsilon^2}\omega_{\rm pl}^2A_{\mu}A'^{\mu}\nonumber\\
 &  +\frac{1}{2}\left(m_A'^2+\frac{\epsilon^2\omega_{\rm pl}^2}{1+\epsilon^2}\right)A'_\mu A'^\mu+\frac{1}{2}\frac{\omega_{\rm pl}^2}{1+\epsilon^2}A^\mu A_\mu
   \label{eq:massbasis_epsilon_medium}
\end{align}
with EOMs in mass basis\footnote{We ignore the QED vacuum polarization effect in appendix for simplicity.} :
\begin{widetext}
\begin{align*}
\begin{cases}
\displaystyle
\left\{\Box + \frac{\omega_{\rm pl}^2}{1+\epsilon^2}\right\}A^\lambda+\kappa  B \partial_l h^{\lambda}-\frac{\epsilon}{1+\epsilon^2} \omega_{\rm pl}^2 A'^\lambda=0,\\
\displaystyle
\left\{\Box + m_{A'}^2+\frac{\epsilon^2}{1+\epsilon^2}\omega_{\rm pl}^2\right\}A'^\lambda+\kappa B'\partial_l h^{\lambda}-\frac{\epsilon}{1+\epsilon^2} \omega_{\rm pl}^2 A^\lambda=0,\\
\Box \tilde{h}^\lambda-\kappa B'\partial_l A'^{\lambda}-\kappa B \partial_l A^\lambda=0.
\end{cases}
\end{align*}
    \end{widetext}
For this set of differential equations, one could easily derive the solution following basically the same procedure as in the main text and find that (again, we expand the result to the leading order of $\mathcal{O}(1/\omega_{\rm pl}^2)$).
To orthogonalize the eigenvectors, we rescale the field vector as

\begin{equation}
    \psi(t_0,r_0)\to\tilde{\psi}(t_0,r_0)\equiv \left\{\sqrt{\tilde{v}_1}A_{\rm I}(t_0,r_0),\sqrt{\tilde{v}_1}A_{\rm S}(t_0,r_0),\tilde{h}(t_0,r_0) \right\},
\end{equation}
with $\tilde{v}_1\equiv \sqrt{1-m_{A'}^2/((1+\epsilon^2)\omega^2)} \approx v+\mathcal{O}(\epsilon^2)$,
and find the eigenvalues and eigenvectors as 
\begin{widetext}
\begin{align}
k_1=\sqrt{\omega^2-\omega_{\rm pl}^2-\frac{m_{A'}^2\epsilon^2}{1+\epsilon^2}},~ \nu_1=\left\{1,~0,~0 \right\}^{\rm T};~~~~~~~~~~~~~~~~~~~~~~~~~~~~~~~~~~~~~~~~\nonumber\\
\begin{cases}
&k_2=\frac{1}{2}\left(\sqrt{\omega^2(1+\tilde{v}_1)^2+(\epsilon B+B')^2\kappa^2/(1+\epsilon^2)}-\sqrt{\omega^2(1-\tilde{v}_1)^2+(\epsilon B+B')^2\kappa^2/(1+\epsilon^2)}\right),\\
&\nu_2=c_2\left\{\displaystyle-\frac{i\epsilon k_2\kappa(\epsilon B+B')}{1+\epsilon^2},~\displaystyle-\frac{i k_2\kappa(\epsilon B+B')}{1+\epsilon^2},~k_2^2-\omega^2\tilde{v}_1^2 \right\}^{\rm T}\nonumber;
\end{cases}\\
\begin{cases}
&k_3=\frac{1}{2}\left(\sqrt{\omega^2(1+\tilde{v}_1)^2+(\epsilon B+B')^2\kappa^2/(1+\epsilon^2)}+\sqrt{\omega^2(1-\tilde{v}_1)^2+(\epsilon B+B')^2\kappa^2/(1+\epsilon^2)}\right),\\
&\nu_3=c_3\left\{\displaystyle-\frac{i\epsilon k_3\kappa(\epsilon B+B')}{1+\epsilon^2},~\displaystyle-\frac{i k_3\kappa(\epsilon B+B')}{1+\epsilon^2},~k_3^2-\omega^2\tilde{v}_1^2 \right\}^{\rm T}\nonumber;
\end{cases}\\
\label{eq:eigens_massbasis}
\end{align}
\end{widetext}
where $c_i$s are again the normalization factors of the eigenvectors. After the replacement of $B,~B'$ 
\begin{align}
\begin{pmatrix}
B\\
B'
\end{pmatrix}=\mathbf{R}
\begin{pmatrix}
\bi\\
\bs
\end{pmatrix},
\label{eq:basistrans_full}
\end{align}
One can clearly see that the eigenvectors $\nu_i$s of different bases could be related by a basis transformation
\begin{align}
    \mathbf{R}'=\begin{pmatrix}
\mathbf{R}&~0\\
0&~1
\end{pmatrix}, 
\end{align}
as anticipated.
The $\mathbf{K}_{3\times3}$ matrix obtained in main-text in flavor basis is 
\begin{align}
    \mathbf{K}_{3\times3}=\begin{pmatrix} 
\sqrt{\omega^2-\omega_{\rm pl}^2} & 0 &0 \\
0 & \frac{v\sqrt{B_{\rm S}^2\kappa^2+(v+1)^2\omega^2}}{v+1}& -\frac{i \kappa B_{\rm S} \sqrt{v}}{v+1} \\
0& \frac{i \kappa B_{\rm S} \sqrt{v}}{v+1}& \frac{\sqrt{B_{\rm S}^2\kappa^2+(v+1)^2\omega^2}}{v+1}\\
\end{pmatrix},
\end{align}

We obtain our final answer for $\mathbf{K}'_{3\times3}$ in the mass basis through Eq.~\eqref{eq:eigens_massbasis}:
\begin{widetext}
\begin{align}
& \mathbf{K}'_{3\times3} =\sum_{i}k_iv_iv_i^\dagger\nonumber\\
&=\begin{pmatrix} 
\sqrt{\omega^2-\omega_{\rm pl}^2+m_{A'}^2(\frac{1}{1+\epsilon^2}-1)}+\frac{\tilde{v}\epsilon^2\sqrt{B_{\rm S}^2\kappa^2+(1+\tilde{v})^2\omega^2}}{(\tilde{v}+1)(1+\epsilon^2)}, & \frac{\tilde{v}\epsilon\sqrt{B_{\rm S}^2\kappa^2+(1+\tilde{v})^2\omega^2}}{(1+\tilde{v})(1+\epsilon^2)} ,&-\frac{iB_{\rm S}\kappa\epsilon \sqrt{\tilde{v}/(1+\epsilon^2)}}{1+\tilde{v}} \\
\frac{\tilde{v}\epsilon\sqrt{B_{\rm S}^2\kappa^2+(1+\tilde{v})^2\omega^2}}{(1+\tilde{v})(1+\epsilon^2)},& \frac{\tilde{v}\sqrt{B_{\rm S}^2\kappa^2+(\tilde{v}+1)^2\omega^2}}{(\tilde{v}+1)(1+\epsilon^2)},& -\frac{i \kappa B_{\rm S} \sqrt{\tilde{v}/(1+\epsilon^2)}}{\tilde{v}+1} \\
\frac{iB_{\rm S}\kappa\epsilon \sqrt{\tilde{v}/(1+\epsilon^2)}}{1+\tilde{v}},& \frac{i \kappa B_{\rm S} \sqrt{\tilde{v}/(1+\epsilon^2)}}{\tilde{v}+1},& \frac{\sqrt{B_{\rm S}^2\kappa^2+(\tilde{v}+1)^2\omega^2}}{\tilde{v}+1}\\
\end{pmatrix}\nonumber\\
&=\mathbf{R}'\cdot \mathbf{K}_{3\times3}\cdot \mathbf{R}'^{\rm T}. ~~~\text{(To leading order of $\epsilon$)}
    \label{eq:homo_res_appd_K}
\end{align}
\end{widetext}

The last equation shows that the independently calculated evolution matrix $\mathbf{K}'$ could be totally related via basis transformations.
Therefore, for a physical process and given different basis calculations we end up with the  same transition probabilities. 

\section{The justification of $\textbf{B}_{\rm S}$ ansatz}
\label{sec:Bs}

Given the spherical symmetry of the solar magnetic field $\bf{B}_{\rm I}$, we can make the $\bf{B}_{\rm S}$ ansatz solution $\mathbf{B}_{\rm S}= b (r) \sin \theta \hat{\phi}$  which exhibits the same spherical symmetry. The $b(r)$ satisfying the radial differential equation \begin{equation}
    \frac{d^2 b(r)}{dr^2}+\frac{2}{r}\frac{d b(r)}{dr^2}- \left(\frac{2}{r^2}+m_{A'}^2 \right)b(r)=-\epsilon m_{A'}^2B_0(r),
\end{equation}
with the boundary condition provided by evaluation of Green function Eq.~\eqref{eq:gr} at specific points is a solution of Eq.~\eqref{eq:bp}.  We next prove that Eq.~\eqref{eq:bp} have only one solution by contradiction that if we have two different solutions $\bf{B}_{\rm S}^1$ and $\bf{B}_{\rm S}^2$  with  $\delta \bf{B}\equiv \bf{B}_{\rm S}^1-\bf{B}_{\rm S}^2$ , we have the Helmholtz equation 
\begin{equation}
    \vec{\nabla}^2 \delta \textbf{B}-m_{A'}^2 \delta \textbf{B}=0
\end{equation}
and the general solutions 
\begin{align}
    &\delta \textbf{B}(r,\theta,\phi)=\nonumber\\
    &\sum_{l=0}^{\infty}\sum_{m=-l}^{+l}\left(\mathbf {a}_{lm}j_l(im_{A'}r)+\mathbf {b}_{lm}y_l(im_{A'}r)\right)Y_l^m(\theta,\phi)
\end{align}
with  $j_l(x)$ and $y_l(x)$ are the spherical Bessel functions,  $Y_{l}^m(\theta,\phi)$ are the spherical harmonics and $\mathbf{a}_{lm},~\mathbf{b}_{lm}$ are the coefficients.  Given the boundary condition that $\delta \textbf{B}=0$ at $r=0, $ $\infty$ and the properties of special functions, we conclude the $\delta \textbf{B}=0$  all the space. Therefore the ansatz solution is the only  solution and the sterile magnetic field exhibit the same spherical symmetry as $\textbf{B}_{\rm I}$.

\bibliography{mainrefs}

\end{document}